\documentclass[11pt,a4paper]{article}
\usepackage{amsmath}

\usepackage{amsfonts,amsmath,amssymb,amsthm}
\usepackage{latexsym,amscd}
\usepackage{amsbsy}
\usepackage{relsize}
\makeatletter

\begin{document}

\title{\bf A Triple-Error-Correcting Cyclic Code from the
    Gold and Kasami-Welch APN Power Functions}

\author{Xiangyong Zeng, Jinyong Shan, Lei Hu
\thanks{X. Zeng and J. Shan are with Faculty of Mathematics and Computer Science, Hubei University, Wuhan 430062,
China (e-mail: xiangyongzeng@yahoo.com.cn).}
\thanks{L. Hu is with the State Key Laboratory of Information Security,
Graduate School of Chinese Academy of Sciences, Beijing 100049,
China (e-mail: hu@is.ac.cn).}}

\maketitle

\begin{quote}
{\small {\bf Abstract:}} Based on a sufficient condition proposed by
Hollmann and Xiang for constructing triple-error-correcting codes,
the minimum distance of a binary cyclic code $\mathcal{C}_{1,3,13}$
with three zeros $\alpha$, $\alpha^3$, and $\alpha^{13}$ of length
$2^m-1$ and the weight divisibility of its dual code are studied,
where $m\geq 5$ is odd and $\alpha$ is a primitive element of the
finite field $\mathbb{F}_{2^m}$. The code $\mathcal{C}_{1,3,13}$ is
proven to have the same weight distribution as the binary
triple-error-correcting primitive BCH code $\mathcal{C}_{1,3,5}$ of
the same length.


\end{quote}

\section{Introduction}

In coding theory, binary triple-error-correcting primitive BCH codes
of length $n=2^m-1$ are one of the most studied objects
\cite{BR,Hoc}. Let $\alpha$ be a primitive element of the finite
field $\mathbb{F}_{2^m}$ with $2^m$ elements, and for a subset $I$ of
$\mathbb{Z}_{2^m-1}$, let $\mathcal{C}_{I}$ denote the length-$n$ cyclic code
with zeros $\alpha^i$ ($i\in I$). The primitive BCH code
$\mathcal{C}_{1,3,5}$ has minimum distance $7$, and its weight
distribution was discussed in \cite{K69, Ber1,Ber2,Ber3}. For some
other integers $d_1$ and $d_2$ (they are naturally assumed to be
different in the sense of cyclotomic equivalence modulo $2^m-1$ and
be different to 1), the code $\mathcal{C}_{1,d_1,d_2}$ can also have
the same weight distribution as the binary triple-error-correcting
primitive BCH code $\mathcal{C}_{1,3,5}$. For example, Table 1 lists all known
such exponent pairs $\{d_1,d_2\}$ for odd $m$, where there exists only one class
of exponents with binary weight greater than $2$, namely
$(2^{\frac{m+1}{2}}+1)^2$ in the construction of \cite{CGHK}.
\begin{table}[htbp]
\caption{ Known exponent pairs $\{d_1, d_2\}$ for odd $m$ such that $\mathcal{C}_{1,d_1,d_2}$ and $\mathcal{C}_{1,3,5}$ have the same
weight distributions }
 \center{\begin{tabular}{|c|c|} \hline $\big\{d_1,d_2\big\}$ &
condition
\\ \hline $\big\{2^r+1,2^{2r}+1\big\}$&$m$ odd, gcd$(m,r)=1$ \cite{K71}
\\ \hline $\big\{2^r+1,2^{3r}+1\big\}$&$m$ odd, gcd$(m,r)=1$ \cite{K71}
\\
\hline $\big\{2^{\frac{m-1}{2}}+1,2^{\frac{m-1}{2}-1}+1\big\}$&$m$ odd \cite{MS}\\
\hline $\big\{2^{\frac{m+1}{2}}+1,(2^{\frac{m+1}{2}}+1)^2\big\}$&$m$ odd \cite{CGHK}\\\hline
\end{tabular}}
\end{table}

Recently, Hollmann and Xiang \cite{HX01} proposed a sufficient
condition for constructing binary triple-error-correcting codes of
length $n=2^m-1$ for odd $m$. More precisely, if a binary cyclic code
$\mathcal{C}$ of length $n=2^m-1$ and dimension $n-3m$ has minimum
distance at least $7$, and if the weights of all codewords of its
dual code $\mathcal{C}^\bot$ are divisible by $2^{\frac{m-1}{2}}$,
then $\mathcal{C}$ has the same weight distribution as the code
$\mathcal{C}_{1,3,5}$. For two exponents $d_1$ and $d_2$ such that
both $x^{d_1}$ and $x^{d_2}$ are almost perfect nonlinear (APN)
power functions from $\mathbb{F}_{2^m}$ to itself, each of the codes
$\mathcal{C}_{1,d_1}$ and $\mathcal{C}_{1,d_2}$ has minimum distance
exactly $5$ by Theorem 5 of \cite{CCZ} (see also Lemma 1 in Section
2). Notice that $\mathcal{C}_{1,d_1,d_2}$ is a subcode of both
$\mathcal{C}_{1,d_1}$  and $\mathcal{C}_{1,d_2}$, then
$\mathcal{C}_{1,d_1,d_2}$ has minimum distance at least $5$. This
motivates us to look for suitable APN power exponents $d_1$ and
$d_2$ such that $\mathcal{C}_{1,d_1,d_2}$ has the same weight
distribution as $\mathcal{C}_{1,3,5}$.

\begin{table}[htbp] \caption{Known values of APN power exponents for odd $m$}
\center{\begin{tabular}{|c|c|c|}
  \hline
  Type & $d$& condition  \\
   \hline Gold & $2^{\,r}+1$ & ${\rm gcd}(r,m)=1$ \cite{G68} \\
  \hline Kasami-Welch & $2^{2r}-2^{r}+1$ & ${\rm gcd}(r,m)=1$ \cite{K71}\\
  \hline Welch  & $2^\frac{m-1}{2}+3$& $m$ odd \cite{Niho} \\
  \hline  Niho & $2^{2r}+2^r-1$ & $4r\equiv-1\,({\rm mod}\,m)$, $m$ odd \cite{Niho} \\
  \hline Inverse & $2^{m-1}-1$ & $m$ odd \cite{BD,N}\\
  \hline Dobbertin & $2^{4r}+2^{3r}+2^{2r}+2^{r}-1$ & $m=5r$, $m$ odd \cite{D3} \\
  \hline
\end{tabular}
}
\end{table}

Following this idea, we experimentally test all known values of APN
power exponents (listed in Table 2) for odd integers $m=5$, 7, 9 and
11, to try to find pairs $(d_1,d_2)$ such that
$\mathcal{C}_{1,d_1,d_2}$ and $\mathcal{C}_{1,3,5}$ have the same
weight distributions. By the MacWilliams identity for binary linear codes \cite{MS}, this is equivalent to say
that their dual codes $\mathcal{C}_{1,d_1,d_2}^\perp$ and
$\mathcal{C}_{1,3,5}^\perp$ have the same weight distributions. The
weight distribution of ${\mathcal{C}}_{1,3,5}^\perp$ is given in
\cite{K69, MS}. The dual code $\mathcal{C}_{1,d_1,d_2}^\bot$ is
simply given by
\begin{equation}\label{DC1}
\begin{array}{c}
\mathcal{C}_{1,d_1,d_2}^\bot={\Big\{}{\bf c}(\epsilon,\gamma,\delta)
=\left({\rm Tr}^m_1(\epsilon x+\gamma x^{d_1}+\delta
x^{d_2})\right)_{x\in \mathbb{F}_{2^m}^*}\mid\epsilon ,\,
\gamma,\,\delta\in \mathbb{F}_{2^m} {\Big\}}
\end{array}
\end{equation}
and its weight distribution is better to compute than that of the target
code $\mathcal{C}_{1,d_1,d_2}$.

All APN exponent pairs $(d_1,d_2)$ such that
${\mathcal{C}}_{1,d_1,d_2}^\perp$ and ${\mathcal{C}}_{1,3,5}^\perp$
have the same weight distributions in our experiment are listed in
Table 3. For odd $m$ and gcd$(r,m)=1$, the code $\mathcal{C}_{2^{r}+1,2^{3r}+1,2^{5r}+1}$ also has the same weight distribution as $\mathcal{C}_{1,3,5}$ \cite{K71}. This construction and those in Table 1 can explain all pairs $\{d_1, d_2\}$ without the mark $\bigstar$ in Table 3. Notice that we say a
pair $(d_1,d_2)$ has actually been explained if $\mathcal{C}_{d,\,2^{i_1}d_1d,\,2^{i_2}d_2d}$ is proven to have the same weight distribution as $\mathcal{C}_{1,3,5}$ for three integers $i_1$, $i_2$, $d$ with $0\leq i_1,i_2\leq m-1$, gcd$(d,2^m-1)=1$ since $\mathcal{C}_{1,d_1,d_2}$ and $\mathcal{C}_{d,\,2^{i_1}d_1d,\,2^{i_2}d_2d}$ have the same weight distributions, where the subscripts are taken modulo $2^m-1$.

\begin{table}[htbp] \caption{Exponent pairs $(d_1, d_2)$  such that
$\mathcal{C}_{1,d_1,d_2}$ and $\mathcal{C}_{1,3,5}$ have the same
weight distributions for $m=5$, $7$, $9$ and $11$ }
\center{\begin{tabular}{|c|c|c|c|c|} \hline Exponent pair
$(d_1,d_2)$ &$m=5$&$m=7$&$m=9$&$m=11$
\\ \hline (Gold, Gold)&(3,5)&(3,5),\,(3,9)&(3,5),\,(3,9)&(3,5),(3,9),(3,17),(3,33)
\\&&(5,9)&(3,17),(5,9)&(5,9),(5,17),(5,33),(9,17)
\\&&&(5,17),(9,17)&(9,33),\,(17,33)
\\
\hline (Gold, Kasami-Welch)&$(3,13)$&$(3,13)^\bigstar$,(9,13)&$(3,13)^\bigstar$&$(3,13)^\bigstar$\\
\hline (Gold, Welch)&(5,7)&(3,11),$(5,11)^{\bigstar}$&&\\
\hline (Gold, Niho)&(3,5)&&&\\
\hline (Kasami-Welch, Welch)&$(13,7)$&&&\\
\hline (Kasami-Welch, Niho)&&$(13,39)$&&\\ \hline
\end{tabular}}
\end{table}

Indeed, we find a new pair marked by $\bigstar$ which can not be explained
by known results, where we regard $(5,11)$ and $(3,13)$ as a same pair since $\mathcal{C}_{1,5,11}$ has the same weight distribution as $\mathcal{C}_{13,\,2\times 5\times 13,\,2^3\times11\times 13}$, i.e., $\mathcal{C}_{1,3,13}$. It is the Gold exponent $d_1=3$ and Kasami-Welch
exponent $d_2=13$, and the latter is another example of exponents
with binary weight $3$.

This paper will prove that for any odd integer $m\geq 5$, the code
$\mathcal{C}_{1,3,13}$ has the same weight distribution as
$\mathcal{C}_{1,3,5}$. To this end, we use a method developed by
Hollmann and Xiang in \cite{HX01,HX02} which analyzes the
divisibility of the weights of the codewords in $\mathcal{C}_{1,3,13}^\perp$
by an add-with-carry algorithm and a technical graph-theoretic deduction. In reference \cite{HX01},
Hollmann and Xiang also applied this method to study the code $\mathcal{C}_{1,d_1,d_2}$ proposed in \cite{CGHK}, where $d_1=2^{\frac{m+1}{2}}+1$ and $d_2=(2^{\frac{m+1}{2}}+1)^2$ are dependent on $m$. The pair $(3,13)$ in this paper is independent on $m$, and this makes the divisibility analysis more complex than that in \cite{HX01}.

The remainder of this paper is organized as follows. Section 2 gives
some preliminaries and the results of this paper. Section 3
establishes a lower bound on the minimum distance of the code
$\mathcal{C}_{1,3,13}$. Section 4 discusses the weight divisibility of
$\mathcal{C}_{1,3,13}^\perp$. Section 5 concludes the study.

\section{Preliminaries and the Results}

Let $\mathbb{F}_{2^m}^*=\mathbb{F}_{2^m}\setminus\{0\}$.
The trace function
${\rm Tr}^m_1$ from $\mathbb{F}_{2^m}$ to $\mathbb{F}_{2}$ is
defined by \cite{LN}
$${\rm Tr}^m_1(x)=\sum\limits_{i=0}^{m-1}
x^{2^{i}},\,\,\,x\in \mathbb{F}_{2^m}.
$$

A binary cyclic code $\mathcal{C}$ of length $n$ is a principal
ideal in the ring $\mathbb{F}_{2}[x]/(x^n-1)$. If  $g(x)$ is a
generator polynomial of $\mathcal{C}$, then a power $\beta$ of a
primitive $n$-th root of unity is a zero of the code  $\mathcal{C}$
if and only if  $g(\beta)=0$. A codeword $c$ in $\mathcal{C}$ has
the form as $c_0+c_1x+\cdots+c_{n-1}x^{n-1}$, which corresponds to a
binary vector $(c_0,c_1,\cdots,c_{n-1})$. The Hamming weight of the
codeword $c$ is the number of nonzero $c_i$ for $0\leq i\leq n-1$,
denoted by $wt(c)$.

{\bf Definition 1:} A function $f$ from $\mathbb{F}_{2^m}$ to itself is said to be almost perfect
nonlinear (APN) if for each $e\in \mathbb{F}_{2^m}^*$, the
function
$\Delta_{f,\,e}(x)=f(x+e)+f(x)$ is two-to-one from $\mathbb{F}_{2^m}$ to itself.

APN functions were introduced in \cite{N} by Nyberg to define them as the
mappings with highest resistance to differential cryptanalysis. For more details we refer
the reader to \cite{BD,BDKM,C,D1,D2,D3,G68,HOU, K71,N}  and the references therein.

For a function $f$ from $\mathbb{F}_{2^m}$ to itself with $f(0)=0$,
let $\mathcal{C}_f$ denote the binary cyclic code of length
$n=2^m-1$ with parity check matrix
$$H_f=\left(
        \begin{array}{ccccc}
          1&\alpha&\alpha^2&\cdots&\alpha^{2^n-2} \\
          f(1)&f(\alpha)&f(\alpha^2)&\cdots&f(\alpha^{2^n-2}) \\
        \end{array}
      \right)
$$
where each entry is viewed as a binary column vector basing on a
basis expression of elements of $\mathbb{F}_{2^m}$ over
$\mathbb{F}_2$.

The APN properties of $f$ can be characterized by the minimum
distance of $\mathcal{C}_f$ \cite{CCZ}.

{\bf Lemma 1:} (\cite{CCZ}) The code $\mathcal{C}_f$ has minimum distance $5$ if
and only if $f$ is APN.

Since the 1960s, the family of triple-error-correcting binary
primitive BCH codes of length $n=2^m-1$ has been thoroughly studied.
The following lemma given by Hollmann and Xiang presented a
sufficient condition for constructing families of
triple-error-correcting codes.

{\bf Lemma 2:} (\cite{HX01}) Let $m$ be odd and $\mathcal{C}$ be a
binary cyclic code of length $n=2^m-1$, dimension $n-3m$ and minimum
distance at least $7$. If all weights of the codewords in $\mathcal{C}^\bot$ are
divisible by $2^{\frac{m-1}{2}}$, then $\mathcal{C}$ has the same
weight distribution as $\mathcal{C}_{1,3,5}$.

With Lemma 2, for odd $m$, we can construct binary
triple-error-correcting codes of length $n=2^m-1$ and dimension
$n-3m$ by analyzing their minimum distances and weight divisibility
of their dual codes. The following Proposition 1 will be proven in the
next section, and the following Lemma 3 shows that the product of
the nonzeros of a binary cyclic code can be used to analyze the
weight divisibility.

{\bf Proposition 1:} For odd $m\geq 5$, the code
$\mathcal{C}_{1,\,3,\,13}$ has minimum distance at least $7$.

{\bf Lemma 3:} (\cite{M71}) Let $\mathcal{C}$ be a binary cyclic
code, and let $l$ be the smallest positive integer such that $l$
nonzeros of $\mathcal{C}$ (with repetitions allowed) have product
$1$. Then the weight of every codeword in $\mathcal{C}$ is divisible
by $2^{l-1}$, and there is at least one codeword whose weight is not
divisible by $2^{l}$.

Based on Lemma 3, Hollmann and Xiang presented an add-with-carry
algorithm to obtain information on the largest power of $2$ dividing
the weights of all codewords of a binary cyclic code as below
\cite{HX01,HX02}.

For a positive integer $m$ and a non-negative integer $a$ with the binary
expression $a=\sum\limits_{i=0}^{m-1}a_i2^i$, $a_i\in \{0,1\}$, the
(binary) weight $w(a)$ of $a$ is defined as the integer
$w(a)=\sum\limits_{i=0}^{m-1}a_i$. For $d_1$, $d_2$, $\cdots$,
$d_j\in \mathbb{Z}_{2^m-1}$, define
$$M(m;d_1, d_2, \cdots, d_j)=\max\left(w(s)-\sum\limits_{l=1}^{j}w(a^{(l)})\right)$$
where the maximum is taken over all integers $s$, $a^{(1)}$,
$\cdots$, $a^{(j)}$ satisfying
$$0\leq s, a^{(1)}, \cdots,
a^{(j)}\leq 2^m-1,\,s\equiv
\sum\limits_{l=1}^{j}d_la^{(l)}\,({\rm mod}\,2^m-1)\,\,{\rm and}\,\,
a^{(l)}\not\equiv 0\,({\rm mod}\,2^m-1)\,{\rm
for\,\, some}\,\, l.$$

The add-with-carry algorithm for integers modulo $2^m-1$ can be
used to determine $M(m;d_1,$ $ d_2,\cdots, d_j)$ \cite{HX01,HX02}.

Let $a^{(l)}$ and $s$ have binary expressions
\begin{equation}\label{INT1}a^{(l)}=\sum\limits_{i=0}^{m-1}
a^{(l)}_i2^i\,\,\,{\rm for}\,\,\,1\leq l\leq j\,\,\,{\rm
and}\,\,\,s=\sum\limits_{i=0}^{m-1}s_i2^i,\end{equation}
respectively. Furthermore, let $d_1$, $d_2$, $\cdots$, $d_j$ be
nonzero integers, and define $d_+=\sum\limits_{d_l>0}d_l$ and
$d_-=\sum\limits_{d_l<0}d_l$ so that
$\sum\limits_{l=1}^jd_l=d_++d_-,\,\,\,
d_+\geq 0,\,\,d_-\leq 0,$ and suppose that
$s\equiv d_1a^{(1)}+d_2a^{(2)}+\cdots+d_ja^{(j)}\,({\rm mod}\,
2^m-1).$

{\bf Lemma 4:} (\cite{HX01,HX02}) There exists a unique integer sequence
$c_{-1},c_0,\ldots,c_{m-1}$ with $c_{-1}=c_{m-1}$ such that
\begin{equation}\label{s:equ1}2c_i+s_i=\sum\limits_{l=1}^jd_la^{(l)}_i+c_{i-1},\quad
0\leq i\leq m-1\end{equation} holds. Moreover, with notation
$w(c)=\sum\limits_{i=0}^{m-1}c_i$, we have that
$$w(c)=\sum\limits_{l=1}^jd_lw(a^{(l)})-w(s).$$ The numbers $c_i$ satisfy
$d_--1\leq c_i\leq d_+$, and further
$$d_-\leq c_i< d_+$$ for all $i$ if $a^{(l)}\not\equiv0\, ({\rm mod}\,
2^m-1)$ holds for some $l$.

The integers $s_i$ and $c_i$ are called the \emph{digits} and
\emph{carries} for the computation of $s$ modulo $2^m-1$ in terms of
$a^{(1)},\cdots,a^{(j)},d_1,\cdots,d_j$.

{\bf Lemma 5:} (\cite{HX01,HX02}) All the weights of
$\mathcal{C}_{1,d_1,d_2}^\bot$ are divisible by
$2^{m-M(m;d_1,d_2)-1}$, and there is at least one codeword whose
weight is not divisible by $2^{m-M(m;d_1,d_2)}$.

The following proposition will be proven in Section 4.

{\bf Proposition 2:} $M(m;3,13)=(m-1)/2$.

By Propositions 1 and 2 and Lemmas 2 and 5, we obtain the following
theorem as the main result in this paper.

{\bf Theorem 1:} For any odd integer $m\geq 5$, the code
$\mathcal{C}_{1,3,13}$ has the same weight distribution as the
binary triple-error-correcting primitive BCH code
$\mathcal{C}_{1,3,5}$.

\section{Minimum Distance of $\mathcal{C}_{1,3,13}$}

{\bf Proof of Proposition 1:} Let $c=(c_0,c_1,\ldots,c_{n-1})$ be an
arbitrary codeword in $\mathcal{C}_{1,\,3,\,13}$, where $n=2^m-1$.
The Discrete Fourier Transform of $c$ is the sequence $\{A_\lambda\}$ with
$$A_\lambda=\sum\limits_{i=0}^{n-1}c_i\alpha^{i\lambda},\,\,\,\,0\leq \lambda<n.$$
From the above formula, we have that $n$ is a period of the sequence $\{A_\lambda\}$. If $A_5=0$, then $c$ is a codeword of the code $\mathcal{C}_{1,3,5}$
which has minimum distance $7$ \cite{K71}. This shows $wt(c)\geq7$.
If $A_9=0$, then
 $c$ is a codeword of the code $\mathcal{C}_{1,3,9}$ which also has minimum distance
$7$ \cite{K71}. Consequently, $wt(c)\geq7$. Thus we can assume that $A_5A_9\neq0$
in the following analysis.

 By \cite{Sch},
the Hamming weight of $c$ equals to the linear complexity (also
called linear span) of the sequence $\{A_\lambda\}$. It is
sufficient to prove that the rank of $M$ is at least $7$, where
\begin{equation}\label{E1}M=\left(\begin{array}{cccc}A_0&A_1&\cdots&A_{n-1}\\
A_1&A_2&\cdots&A_{0}\\
\vdots&\vdots&&\vdots\\
A_{n-1}&A_0&\cdots&A_{n-2}\end{array}\right).\end{equation}
To this end, we will argue separately according to the
parity of  $wt(c)$.

(1) Suppose that $wt(c)$ is odd, i.e., $A_0=1$.

In this case, we will find two submatrices $M_1$ and $M_2$ of $M$
such that either $M_1$ or $M_2$ has full rank, where
$$M_1=\begin{pmatrix}A_0&A_1&A_2&A_4&A_6&A_8\\
A_1&A_2&A_3&A_5&A_7&A_9\\
A_2&A_3&A_4&A_6&A_8&A_{10}\\
A_3&A_4&A_5&A_7&A_9&A_{11}\\
A_5&A_6&A_7&A_9&A_{11}&A_{13}\\
A_6&A_7&A_8&A_{10}&A_{12}&A_{14}\end{pmatrix}\quad {\rm and }\quad
M_2=\begin{pmatrix}A_0&A_1&A_3&A_4&A_7&A_8\\
A_1&A_2&A_4&A_5&A_8&A_9\\
A_2&A_3&A_5&A_6&A_9&A_{10}\\
A_3&A_4&A_6&A_7&A_{10}&A_{11}\\
A_4&A_5&A_7&A_8&A_{11}&A_{12}\\
A_5&A_6&A_8&A_{9}&A_{12}&A_{13}\end{pmatrix}.$$

Notice that $A_{\lambda}=0$ if $\lambda\in C_1\cup C_3\cup C_{13}$, where $C_i$ denotes the cyclotomic coset  modulo $2^m-1$ containing the integer $i$.
Consequently, we have $A_1=A_2=A_3=A_4=A_6=A_8=A_{12}=A_{13}=0$.
From the expression of $A_\lambda$, we have $A_{10}=A_5^2$,
$A_{14}=A_7^2$ and $A_{18}=A_9^2$.

It can be directly verified that
$$\det(M_1)=A_5^2A_7(A_7^3+A_5^2A_{11})\,\,\,\,{\rm and}\,\,\,\,
\det(M_2)=A_5^2(A_5^2A_9^2+A_5A_9A_7^2+A_5^2A_7A_{11}).$$ If
$A_7=0$, then $\det(M_2)=A_5^4A_9^2\neq0$ by our assumption that
$A_5A_9\neq0$, i.e., rank$(M_2)=6$. If $A_7\neq0$ and $A_{11}=0$,
then $\det(M_1)\neq0$ by  $A_5A_7\neq0$, i.e., $M_1$ has rank $6$.
If $A_7\neq0$, $A_{11}\neq0$ and $\det(M_1)=0$, then
$A_7^3=A_5^2A_{11}$. Thus,
\begin{equation}\label{E2}\det(M_2)=A_5^2(A_5^2A_9^2+A_5A_9A_7^2+A_7^4),\end{equation}
which is either $A_5^3A_9A_7^2\neq0$ if $A_5A_9=A_7^2$ or
$$A_5^2(A_5A_9+A_7^2)^{-1}\Big[(A_5A_9)^3+(A_7^2)^3\Big]\neq0$$
since $\gcd (3,n)=1$ if $A_5A_9\neq A_7^2$. Therefore, either $M_1$
or $M_2$ has full rank, and then rank$(M)\geq 6$. As a consequence,
$wt(c)\geq 7$.

(2) Suppose that $wt(c)$ is even, i.e., $A_0=0$.

If $A_7=0$, we will prove the following submatrix
\[M_3=\begin{pmatrix}A_0&A_1&A_2&A_4&A_5&A_6&A_8\\
A_1&A_2&A_3&A_5&A_6&A_7&A_9\\
A_2&A_3&A_4&A_6&A_7&A_8&A_{10}\\
A_4&A_5&A_6&A_8&A_{9}&A_{10}&A_{12}\\
A_5&A_6&A_7&A_{9}&A_{10}&A_{11}&A_{13}\\
A_7&A_8&A_9&A_{11}&A_{12}&A_{13}&A_{15}\\
A_8&A_{9}&A_{10}&A_{12}&A_{13}&A_{14}&A_{16}\end{pmatrix}\] has rank
$7$. By a direct calculation, we have $\det(M_3)=A_5^7A_9^2\neq0$.
Thus rank$(M_3)\geq7$ which implies that $wt(c)\geq7$.

If $A_7\neq0$, we will prove the submatrix
$$M_4=\begin{pmatrix}A_0&A_1&A_2&A_4&A_6&A_7&A_8\\
A_1&A_2&A_3&A_5&A_7&A_8&A_9\\
A_2&A_3&A_4&A_6&A_8&A_9&A_{10}\\
A_4&A_5&A_6&A_8&A_{10}&A_{11}&A_{12}\\
A_5&A_6&A_7&A_9&A_{11}&A_{12}&A_{13}\\
A_8&A_9&A_{10}&A_{12}&A_{14}&A_{15}&A_{16}\\
A_{12}&A_{13}&A_{14}&A_{16}&A_{18}&A_{19}&A_{20}\end{pmatrix}$$ has
rank $7$. By a direct calculation, we have
$\det(M_4)=A_5^5A_7(A_5^2A_9^2+A_5A_9A_7^2+A_7^4)$. With a similar
analysis as for (\ref{E2}), we have $\det(M_4)\neq0$ and then
rank$(M)\geq 7$. Thus, $wt(c)\geq7$.  \quad$\blacksquare$

{\bf Remark 1:} The reference \cite{Sch} showed that the minimum distance of a
linear cyclic code is equal to the rank of a matrix constructed by using Discrete Fourier Transform. This together with
BCH or HT bound established  a lower bound on the minimum distance
of the code proposed in \cite{CGHK}. In Proposition 1, we apply this
method and the results for the minimum distances of the cyclic codes
$\mathcal{C}_{1,3,5}$ and $\mathcal{C}_{1,3,9}$ \cite{K71} to obtain
a lower bound on minimum distance of $\mathcal{C}_{1,3,13}$.

\section{Divisibility of Weights in $\mathcal{C}_{1,3,13}^\bot$}

In this section, for an odd integer $m=2k+1$ with $k\geq 2$, we
will prove $M(m;3,13)= k$.

Let $s$, $a$ and $b$ be integers with $0\leq s$, $a$, $b\leq 2^m-1$,
$s\equiv 3a+13b\,({\rm mod}\, 2^m-1)$, and assume that at least one
of $a$ and $b$ is nonzero modulo $2^m-1$. Let
$s=\sum\limits_{i=0}^{m-1}s_i2^i$,
$a=\sum\limits_{i=0}^{m-1}a_i2^i$, and
$b=\sum\limits_{i=0}^{m-1}b_i2^i$ be the binary expressions of $s$,
$a$ and $b$, respectively.

We first prove  $M(m;3,13)\leq k$, namely $w(s)-w(a)-w(b)\leq k$ in
the sequel.

Notice that $2a$, $8b$,
$4b\,({\rm mod}\,2^m-1)$  have the binary expressions
$\sum\limits_{i=0}^{m-1}a_{i-1}2^i$,
$\sum\limits_{i=0}^{m-1}b_{i-3}2^i$,
$\sum\limits_{i=0}^{m-1}b_{i-2}2^i$, respectively, and $s\equiv
3a+13b\equiv 2a+a+8b+4b+b\,({\rm mod}\, 2^m-1)$. Taking $d_l=1$ for $l\in\{1,2,3,4,5\}$ and $a^{(1)}=2a$, $a^{(2)}=a$, $a^{(3)}=8b$, $a^{(4)}=4b$, $a^{(5)}=b$ and applying Lemma 4,  there are carries $c_i\in \{0,1,2,3,4\}$ such that
\begin{equation}\label{Cr1}2c_i+s_i=a_{i-1}+a_i+b_{i-3}+b_{i-2}+b_i+c_{i-1},\,\,\,0\leq i\leq m-1,\end{equation}
where the subscripts are taken modulo $m$. With
$w(c)=\sum\limits_{i=0}^{m-1}c_i$, by the $m$ equalities in
(\ref{Cr1}) we have
\begin{equation}\label{Cr2}w(c)+w(s)=2w(a)+3w(b).\end{equation}  Let
\begin{equation}\label{Cr3}\nu_i=a_{i-1}+a_i+b_{i-3}+b_{i-2}+b_{i-1}+b_i-c_{i-1}-c_i,\,\,\,0\leq i\leq m-1\end{equation}
and  $w(\nu)=\sum\limits_{i=0}^{m-1}\nu_i$. Then by (\ref{Cr3}) and (\ref{Cr2}), we have
\begin{equation}\label{Cr4}w(\nu)=2w(a)+4w(b)-2w(c)=2\big(w(s)-w(a)-w(b)\big).\end{equation}

To prove $w(s)-w(a)-w(b)\leq k$, by (\ref{Cr4}) it is sufficient
to prove $w(\nu)\leq m$. To this end, we will define a certain
weighted directed graph $\mathbb{D}$ and recall some related definitions in \cite{BM} as below.

A \emph{directed graph} $\mathbb{D}$ is an ordered pair
$(V(\mathbb{D}),A(\mathbb{D}))$ consisting of a set $V(\mathbb{D})$
of vertices and a set $A(\mathbb{D})$, disjoint from
$V(\mathbb{D})$, of \emph{arcs}, together with an \emph{incidence
function} $\psi_\mathbb{D}$ that associates with each arc
$\vartheta$ of $\mathbb{D}$ an ordered pair of (not necessarily
distinct) vertices
$\psi_\mathbb{D}(\vartheta)=(T(\vartheta),H(\vartheta))$ of
$\mathbb{D}$. The vertex $T(\vartheta)$ is the {\it tail} of
$\vartheta$, and the vertex $H(\vartheta)$ its \emph{head}. For each
arc $\vartheta$ in a directed graph $\mathbb{D}$, we can associate a
real number $w(\vartheta)$ with $\vartheta$, and $w(\vartheta)$ is
called its \emph{weight}. In this case, $\mathbb{D}$ is called to be
a {\it weighted directed graph}. In a directed graph $\mathbb{D}$, a
\emph{directed walk} is an alternating sequence of vertices and arcs
\[W:=P_0\vartheta_0 P_1\cdots P_{l-1}\vartheta_{l-1}P_l\] such that
for each $i$ with $1\leq i\leq l$, $P_{i-1}$ and $P_i$ are the tail
and head of $\vartheta_{i-1}$, respectively. In this case, we refer
to $W$ as a \emph{directed} $(P_0,P_l)$\emph{-walk}. For two
vertices $P_i$ and $P_j$ in the walk $W$ where $0\leq i<j\leq l$,
the $(P_i,P_j)$-\emph{segment} of $W$ is the subsequence of $W$
starting with $P_i$ and ending with $P_j$, and it is denoted
$P_iWP_j$. The directed walk $W$ in $\mathbb{D}$ is \emph{closed} if
its initial and terminal vertices $P_0$, $P_l$ are identical.

With these preparations, we can define a weighted directed graph
$\mathbb{D}$. The vertices of $\mathbb{D}$ consist of all vectors
$P=(x,y,z,u)$, where $x,y,z\in \{0,1\}$ and $u\in \{0,1,2,3,4\}$.
Let $P_1=(x_1,y_1,z_1,u_1)$ and $P_2=(x_2,y_2,z_2,u_2)$ be two
vertices of $\mathbb{D}$, and define an arc $\vartheta$ with
$T(\vartheta)=P_1$ and $H(\vartheta)=P_2$ if
\begin{equation}\label{Con1}x_1+y_1+z_1+x_2+z_2-2u_1+u_2=0,\, \, {\rm or}\, \,1.\end{equation}
The weight of the arc $\vartheta$ is defined as
$$w(\vartheta)=x_1+y_1+z_1+x_2+y_2+z_2-u_1-u_2.$$ Thus for
$i\in\{0,1,\cdots,m-1\}$,
\begin{equation}\label{vertice}V_i=(a_i,b_i,b_{i-2},c_i)\end{equation}
are $m$ vertices of $\mathbb{D}$, where $a_i$, $b_i$, and  $c_i$ are
those integers in (\ref{Cr1}). Furthermore, there are $m$ arcs
$\vartheta_i$ with $w(\vartheta_i)=\nu_{i}$ defined by (\ref{Cr3})
with the tail  $V_i=(a_i,b_i,b_{i-2},c_i)$ and head
$V_{i-1}=(a_{i-1},b_{i-1},b_{i-3},c_{i-1})$ for all $0\leq i\leq
m-1$ since
$a_i+b_i+b_{i-2}+a_{i-1}+b_{i-3}-2c_i+c_{i-1}=s_i\in\{0,1\}$ by
(\ref{Cr1}), where the subscripts are taken modulo $m$.

With the help of a computer, we have that there are totally $320$
arcs in $\mathbb{D}$, and their weight distribution is given in
Table 4. Furthermore, every vertex in the set
\begin{equation}\label{Set1}\Gamma=\Big\{(1,1,0,0), (1,0,1,0),
(0,1,1,0), (1,1,1,0)\Big\}\end{equation} cannot be the tail of any arc
in $\mathbb{D}$. Some arcs $\vartheta$ with head $H(\vartheta)\not\in\Gamma$ will
be used in this section and they are listed in Appendix A.

\begin{table}[htbp] \caption{The weight distribution of all arcs in the
weighted directed graph $\mathbb{D}$}
\center{\begin{tabular}{|c|c|c|c|c|c|c|c|c|c|c|c|} \hline
Weight&-6&-5&-4&-3&-2&-1&0&1&2&3&4\\ \hline The number of arcs&1&16&36&43&43&42&43&43&36&16&1\\
\hline
\end{tabular}}
\end{table}

Notice that for the case $\nu_i<2$ for all $i\in\{0,1,\cdots,m-1\}$,
it can be easily verified that $w(\nu)\leq m$. Consequently, the
proof for $w(\nu)\leq m$ can be proceeded in two steps as below.

{\bf Step 1}: To prove that for any $\nu_i\geq 2$, there exists a positive
integer $t\leq m$ such that
$\nu_i+\nu_{i-1}+\cdots+\nu_{i-t+1}\leq  t$.

{\bf Step 2}: Based on Step 1, we will prove $w(\nu)=\sum\limits_{i=0}^{m-1}\nu_i\leq m$.

The two steps are summarized as the following Propositions 3 and 4.

{\bf Proposition 3:} For any $\nu_i\geq 2$, there exists a positive
integer $t\leq m$ such that $\nu_i+\nu_{i-1}+\cdots+\nu_{i-t+1}\leq
t$, where the subscripts are taken modulo $m$.

By the weighted directed graph $\mathbb{D}$ defined as above, the
number $\nu_i+\nu_{i-1}+\cdots+\nu_{i-t+1}$ can be regarded as the
sum of the weights of some arcs in $\mathbb{D}$. To finish the proof
of Proposition 3, we need to study a set
\begin{equation}\label{P}{\mathcal{P}} =\Big\{W= P_0\vartheta_0 P_1 \vartheta_1 \cdots
P_{i-1}\vartheta_{i-1}P_i\vartheta_i\cdots
P_{q-1}\vartheta_{q-1}P_{q}\,|\,W\in {\mathbb{D}},\,q{\rm \,is\,dependent\, on} \,W\Big\}\end{equation}
consisting of all directed walks $W$ with the following properties:
\begin{enumerate}
\item [(I)] any vertex of the set $\Gamma$ in (\ref{Set1}) does not
occur in $W$;
\item [(II)] for $0\leq i \leq q-2$, any three consecutive vertices
$P_{i}$, $P_{i+1}$, and $P_{i+2}$ in $W$ satisfy
$P_{i}(3)=P_{i+2}(2)$, where $P_i(l)$ denotes the $l$-th  component
of $P_{i}$ for $l\in\{1,2,3,4\}$; in addition, if the walk $W$ is
closed, then  $P_{q-1}(3)=P_1(2)$;
\item [(III)] any arc $\vartheta_i$ in $W$ satisfies that
$w(\vartheta_i)\geq (i+2)-T_i$ for $0\leq i\leq q-1$, where
$T_0=0$ and $T_i=\sum\limits_{l=0}^{i-1}w(\vartheta_l)$ for $i\geq 1$.
\end{enumerate}

If Proposition 3 cannot be true, then there is an integer $i_0$
with $0\leq i_0\leq m-1$ such that $\nu_{i_0}\geq 2$ and
$\nu_{i_0}+\nu_{{i_0}-1}+\cdots+\nu_{{i_0}-t+1}\geq t+1$ for any
positive integer $t$ with $2\leq t\leq m$. Let
\begin{equation}\label{W0} W_0=P_0\vartheta_0 P_1 \vartheta_1 \cdots
P_{i-1}\vartheta_{i-1}P_{i}\vartheta_i\cdots
P_{m-2}\vartheta_{m-2}P_{m-1}\end{equation} be the walk such that
$P_i=V_{i_0-i}$ in (\ref{vertice}) for $0\leq i\leq m-1$, and
$\vartheta_i$ be the arc with $T(\vartheta_i)=P_i$ and
$H(\vartheta_i)=P_{i+1}$ for $i\in\{0,1,\cdots,m-1\}$, where the
subscripts are taken modulo $m$. Then, we have $w(\vartheta_0)\geq
2$ and for any positive integer $t$ with $2\leq t\leq m$ such that
$w(\vartheta_0)+w(\vartheta_1)+\cdots+w(\vartheta_{t-1})\geq t+1$.
Thus by (\ref{vertice}) and the analysis therein,  $W_0\in
\mathcal{P}$ and it is closed. As a consequence, it will lead to a
contradiction if any walk $W\in \mathcal{P}$ is not closed. In fact,
we can prove that any walk $W\in \mathcal{P}$ is not closed in the
sequel. This will give the proof of Proposition 3.

The following notations are
used throughout this section:

\begin{enumerate}
\item [$\bullet$]
 $P_i\xrightarrow[]{(\eta,\,\omega)}$ denotes any walk $P_i\vartheta_iP_{i+1}$
 with $T(\vartheta_i)=P_i$, $H(\vartheta_i)=P_{i+1}$,
$P_{i+1}(2)=\eta$ and $w(\vartheta_i)\geq \omega$;
\item [$\bullet$] $P_i\xrightarrow[]{(-,\,\omega)}$  denotes any walk
$P_i\vartheta_iP_{i+1}$ with $T(\vartheta_i)=P_i$, $H(\vartheta_i)=P_{i+1}$,
$P_{i+1}(2)\in\{0,1\}$ and $w(\vartheta_i)\geq
\omega$;
\item [$\bullet$] $P_i\xrightarrow[]{(\eta,\,\omega)}O$ denotes that there does
not exist any arc $\vartheta$ such that $T(\vartheta)=P_i$, $H(\vartheta)\in\mathbb{D}$,
$(H(\vartheta))(2)=\eta$ and
$w(\vartheta)\geq
\omega$.
\end{enumerate}

With the above notations, we can conveniently describe the walks in $\mathcal{P}$.

{\bf Example 1:} Let $q$ be a positive integer and
$\omega=(j+2)-T_{j}=1$ for some positive integer $j$ with $0\leq
j<q$, and let $$W:\,\,\,P_0 \xrightarrow[]{} P_1
\xrightarrow[]{}\cdots
\xrightarrow[]{} P_{j-1}\xrightarrow{}P_j=(0,0,0,0)\xrightarrow[]{(0,\,\omega)} P_{j+1}
\xrightarrow[]{}P_{j+2} \xrightarrow[]{}\cdots
\xrightarrow[]{}P_{q}$$ be a walk in the set $\mathcal{P}$, and
$\vartheta_i$ be the arc with the tail $P_i$ and head $P_{i+1}$ for
each $i\in\{0,1,\cdots,q-1\}$. By Appendix A, we can find all
possibilities for the segment $P_{j+1}WP_q$, which is completely
determined by the walk $(0,0,0,0)\xrightarrow[]{(0,\,1)}$.

If we find all possibilities for the segment $P_{j+1}WP_q$, then we
also know all possibilities for the segment $P_{j+1}WP_{q'}$ for any
integer $j+1\leq q'\leq q$. Therefore, without loss of generality,
we can assume that the integer  $q$ is large enough.

Since $P_{j+1}(2)=0$ and $w(\vartheta_j)\geq 1$, by Appendix A,
we have $P_{j+1}\in \big\{(1,0,0,0),(0,0,1,0)\big\}$. If $P_{j+1}=(1,0,0,0)$,
 by Properties (II) and (III) of the walks in
$\mathcal{P}$, we have $P_{j+2}(2)=P_j(3)=0$ and
$w(\vartheta_{j+1})\geq
(j+3)-T_{j+1}=(j+3)-w(\vartheta_j)-T_j=(j+2)-w(\vartheta_j)-T_j=1$.
By Appendix A, we can uniquely determine $P_{j+2}=(0,0,0,0)$.
Furthermore, with $w(\vartheta_{j+1})=1$ and $P_{j+1}=(1,0,0,0)$, we
have
\begin{equation}\label{INQ3}w(\vartheta_{j+2})\geq (j+4)-T_{j+2}=(j+4)-w(\vartheta_{j+1})-T_{j+1}=(j+3)-T_{j+1}=1\end{equation}
and $P_{j+3}(2)=0$. Therefore, for $P_{j+1}=(1,0,0,0)$, $P_{j}WP_{j+3}$ can be expressed as
\begin{equation}\label{Chasing1}
(0,0,0,0)\xrightarrow{(0,1)}(1,0,0,0)\xrightarrow{(0,\,1)}(0,0,0,0)\xrightarrow{(0,\,1)}.
\end{equation}
Similarly, for $P_{j+1}=(0,0,1,0)$,
$P_{j}WP_{j+5}$ is given by
\begin{equation}\label{Chasing2}
(0,0,0,0)\xrightarrow{(0,1)}(0,0,1,0)\xrightarrow{(0,\,1)}(0,0,0,0)\xrightarrow{(1,\,1)}(0,1,0,0)
\xrightarrow{(0,\,1)}(0,0,0,0)\xrightarrow{(0,\,1)}.
\end{equation}
Combining (\ref{Chasing1}) and (\ref{Chasing2}), we have an expression consisting of two segments with initial vertex $P_j$
\begin{equation}\label{CHASE}
(0,0,0,0)\xrightarrow{(0,1)}\left\{
\begin{array}{l}(1,0,0,0)\xrightarrow{(0,\,1)}(0,0,0,0)\xrightarrow{(0,\,1)}\\[7pt]
(0,0,1,0)\xrightarrow{(0,\,1)}(0,0,0,0)\xrightarrow{(1,\,1)}(0,1,0,0)\xrightarrow{(0,\,1)}(0,0,0,0)\xrightarrow{(0,\,1)}.
\end{array}\right.
\end{equation}
In the first segment of (\ref{CHASE}), $P_{j+3}=(1,0,0,0)$ or
$(0,0,1,0)$ since $(0,0,0,0)\xrightarrow{(0,1)}$ has only two
possible forms,  which have occurred as $P_jWP_{j+1}$ in  the first and
second  segments of (\ref{CHASE}), respectively. By a similar analysis,
we have $P_{j+5}=(1,0,0,0)$ or $(0,0,1,0)$ in the second segment of
(\ref{CHASE}). Therefore, again by (\ref{CHASE}), we have that
$P_{j+3}WP_{j+5}$ has the form as
\begin{equation}\label{Chasing21}(1,0,0,0)\xrightarrow{(0,\,1)}(0,0,0,0)\xrightarrow{(0,\,1)}\end{equation}
or $P_{j+3}WP_{j+7}$ has the form as
\begin{equation}\label{Chasing22}(0,0,1,0)\xrightarrow{(0,\,1)}(0,0,0,0)\xrightarrow{(1,\,1)}(0,1,0,0)\xrightarrow{(0,\,1)}
(0,0,0,0)\xrightarrow{(0,\,1)}\end{equation} in the first segment of
(\ref{CHASE}). Similarly, we have that $P_{j+5}WP_{j+7}$ has the
form as  (\ref{Chasing21}) or $P_{j+5}WP_{j+9}$ has the form as
(\ref{Chasing22}) in the second segment of (\ref{CHASE}). Repeating
the above process, all possibilities of $P_{j+1}WP_q$ can be
obtained. Further, all vertices $P_l$ $(j\leq l\leq q)$ have occurred
in the two segments of (\ref{CHASE}), and they are $(0,0,0,0)$,
$(1,0,0,0)$, $(0,1,0,0)$, and $(0,0,1,0)$.

\textbf{Remark 2:} In Example 1,  $(0,0,0,0)\xrightarrow{(0,1)}$
completely determines all possibilities for the segment
$P_{j+1}WP_q$ of $W$. The expression (\ref{CHASE}) consists of two
basic segments of $W$, by which all possibilities of the segment
$P_{j+1}WP_q$ can be conveniently found. In the proofs of Lemmas 6
and 7, for some given $P_j\xrightarrow{(\eta,\omega)}$ of a walk $W$
in $\mathcal{P}$,  we will frequently need to determine all
possibilities for the segment $P_{j+1}WP_q$ of $W$. Similarly as in
Example 1, we will use some expression consisting of basic segments
of $W$ to determine all possibilities of $P_{j+1}WP_q$. We call the
expression as (\ref{CHASE}) {\it a set of basic segments} (SBS) of
$P_j\xrightarrow{(\eta,\omega)}$.

The following two lemmas will be used to prove Proposition 3.

{\bf Lemma 6:} Let $q$ be a positive integer and
$\omega=(j+2)-T_{j}$ for some positive integer $j$ with $0\leq j<q$.
 For any walk
\[W:\,\,\,P_0 \xrightarrow[]{}P_1 \xrightarrow[]{} \cdots
\xrightarrow[]{}P_{j-1}\xrightarrow[]{}P_j=(0,0,0,0)\xrightarrow[]{(-,\,\omega)}
P_{j+1}
\xrightarrow[]{}\cdots\xrightarrow[]{}
P_{q-1}\xrightarrow[]{}P_{q}\]
in the set $\mathcal{P}$ defined by (\ref{P}), we have

\noindent(i) if $\omega=0$ or $1$, all vertices $P_{l}$ $(j+1\leq l\leq q)$
occurring in the walk $W$ are contained in the set
\begin{equation}\label{LCL:eq4}\begin{array}{r}
S_1=\Big\{(0,0,0,0),(0,0,1,0),(0,1,0,0),(1,0,0,0),(0,1,0,1)\Big\};
\end{array}
\end{equation}

\noindent(ii) if $\omega=-1$, all vertices $P_{l}$ $(j+1\leq l\leq q)$
occurring in the walk $W$ are contained in the set
\begin{equation}\label{LCL:eq3}\begin{array}{l}
S_2=S_1\bigcup\Big\{(0,0,0,1), (0,0,1,1),(1,0,0,1)\Big\};
\end{array}
\end{equation}

\noindent(iii) if $\omega=-2$, all vertices $P_{l}$ $(j+1\leq l\leq q)$
occurring in the walk $W$ are contained in the set
\begin{equation}\label{LCL:eq2}\begin{array}{l}
S_3=S_2\bigcup\Big\{(1,0,1,1),(0,1,1,1),(1,1,0,1)\Big\}.
\end{array}
\end{equation}

The proof of Lemma 6 is presented in Appendix B.

{\bf Lemma 7:} For the walk
\[W:\,\,\,P_0\xrightarrow[]{} P_1 \xrightarrow[]{}
\cdots
\xrightarrow[]{}P_{q-1}\xrightarrow[]{}P_q\]
in the set $\mathcal{P}$, if the initial vertex $P_0\in \big\{(1,0,0,0), (0,1,0,0), (0,0,1,0),
(1,0,1,1), (1,1,0,1)$, $(0,1,1,1)\big\}$, then $W$ cannot be closed.

{\bf Proof:} Let $\vartheta_j$ denote the arc with the tail $P_j$ and head $P_{j+1}$ for each $j\in\{0,1,\cdots,q-1\}$. Since $W\in \mathcal{P}$, by Property (III) of the walks in $\mathcal{P}$, we have $w(\vartheta_0)\geq 2$. If $W$ is closed, then we must have $P_q=P_0$ and $P_{q-1}(3)=P_1(2)$. The lemma is proven according to six cases of the vertex $P_0$ as follows.

If $P_0=(1,0,0,0)$ and $w(\vartheta_0)\geq 2$, then
$P_1=(0,1,0,0)$ by Appendix A. Consequently, $P_2(2)=0$ and  by Property (III) of the walks in $\mathcal{P}$,
$w(\vartheta_1)\geq 1$. By a similar analysis as in Example 1, $(0,1,0,0)\xrightarrow{(0,1)}$ has an SBS as
\begin{equation}\label{NEW6}(0,1,0,0)\xrightarrow{(0,1)}
(0,0,0,0)\xrightarrow{(0,1)}\left\{\begin{array}{l}
(1,0,0,0)\xrightarrow{(0,1)}(0,0,0,0)\xrightarrow{(0,1)}\\
(0,0,1,0)\xrightarrow{(0,1)}(0,0,0,0)\xrightarrow{(1,1)}(0,1,0,0)\xrightarrow{(0,1)}.\end{array}\right.\end{equation}
From (\ref{NEW6}), we can know that all vertices and arcs in
$P_1WP_q$ have occurred in (\ref{NEW6}). If $P_q=P_0=(1,0,0,0)$,
then by (\ref{NEW6}), $P_{q-1}=(0,0,0,0)$ and then $P_{q-1}(3)=0\neq
P_1(2)$. Therefore the walk $W$ cannot be closed if $P_0=(1,0,0,0)$.

The case $P_0=(0,1,0,0)$ can be similarly proven as the case $P_0=(1,0,0,0)$.

If $P_0=(0,0,1,0)$, then $P_0WP_4$ has the form as
\begin{equation}\label{l:equ11}(0,0,1,0)\xrightarrow{(-,2)}(0,1,0,0)
\xrightarrow{(1,1)}(0,1,0,0)\xrightarrow{(0,0)}(0,0,0,0)\xrightarrow{(0,0)}.
\end{equation}
If $W$ is closed, then $P_q=(0,0,1,0)$ and $P_{q-1}(3)=1$. By
(\ref{l:equ11}), we have $q\geq 5$. By Lemma 6 (i),  the vertices
$P_j$ for $4\leq j\leq q$ in $W$ are contained in $S_1$.
Consequently, $P_{q-1}\in S_1$. Notice that $(0,0,1,0)$ is the
unique vertex with the third component $1$ in the set $S_1$. As a
consequence, $P_{q-1}=(0,0,1,0)$ and the arc $\vartheta_{q-1}$ is
$(0,0,1,0)\xrightarrow[]{}(0,0,1,0)$, which does not exist by
Appendix A. This leads to a contradiction and then $W$ cannot be
closed.

If $P_0=(1,0,1,1)$, then $(1,0,1,1)\xrightarrow[]{(-,2)}$ has an SBS as
\begin{equation}\label{l:equ15}(1,0,1,1)\xrightarrow[]{(-,2)}\left\{\begin{array}{l}
(1,0,0,0)\xrightarrow{(1,1)}(0,1,0,0)\xrightarrow{(0,0)}(0,0,0,0)\xrightarrow{(0,0)}\\
(0,1,0,0)\xrightarrow{(1,1)}(0,1,0,0)\xrightarrow{(0,0)}(0,0,0,0)\xrightarrow{(0,0)}\\
(0,0,1,0)\xrightarrow{(1,1)}(0,1,0,0)\xrightarrow{(1,0)}(0,1,0,0)\xrightarrow{(0,-1)}(0,0,0,0)\xrightarrow{(0,-1)}.
\end{array}\right.\end{equation} The vertices $P_j$ for $4\leq j\leq q$
of the first and second segments of (\ref{l:equ15}) are contained in
$S_1$ and the vertices $P_j$ for $5\leq j\leq q$ of the third
segment in (\ref{l:equ15}) are contained in $S_2$ by Lemma 6 (i) and
(ii). Notice that $(1,0,1,1)\not\in S_1$ and $(1,0,1,1)\not\in S_2$.
Consequently, the walk $W$ cannot be closed.

If $P_0=(1,1,0,1)$, then $P_0WP_3$ has three possible forms as
\begin{equation*}\label{equ1}(1,1,0,1)\xrightarrow[]{(-,2)}\left\{\begin{array}{l}
(1,0,0,0)\xrightarrow{(0,1)}(0,0,0,0)\xrightarrow{(0,1)}\\
(0,1,0,0)\xrightarrow{(0,1)}(0,0,0,0)\xrightarrow{(0,1)}\\
(0,0,1,0)\xrightarrow{(0,1)}(0,0,0,0)\xrightarrow{(1,1)}.
\end{array}\right.\end{equation*} The vertices $P_j$ for $3\leq j\leq q$ are contained in $S_1$ by Lemma
6 (i). The fact $(1,1,0,1)\not\in S_1$ implies that $W$ cannot be closed.

The case $P_0=(0,1,1,1)$ can be similarly proven as the
case $P_0=(1,0,1,1)$.

The proof is finished.\quad$\blacksquare$

Applying Lemmas 6 and 7, we will finish the proof of Proposition 3
as below.

{\bf Proof of Proposition 3:} If the result is not true, the walk
$W_0$ defined in (\ref{W0}) belongs to the set $\mathcal{P}$ and
$w(\vartheta_0)\geq 2$. We will prove that $W_0$ cannot be closed
according to $\vartheta_0$.

Notice that there are no arcs $\vartheta$ with tail $T(\vartheta)\in
\Gamma$, where $\Gamma$ is defined by (\ref{Set1}). As a
consequence, $W_0$ cannot be closed if $\vartheta_0$ occurs in Table
5.

\begin{table}[htbp] \caption{All arcs $\vartheta$ with $w(\vartheta)\geq 2$ and $H(\vartheta)\in \Gamma$}
\center{\begin{tabular}{|c|c|c|c|c|c|c|c|c|} \hline $T(\vartheta)$&$H(\vartheta)$&$w(\vartheta)$&$T(\vartheta)$&$H(\vartheta)$&$w(\vartheta)$&$T(\vartheta)$&$H(\vartheta)$&$w(\vartheta)$
\\ \hline (0,0,0,0)&(1,1,0,0)&2&
(0,0,0,0)&(0,1,1,0)&2&(0,0,0,1)&(1,1,1,0)&2\\ \hline
(1,0,0,1)&(1,1,0,0)&2& (1,0,0,1)&(1,0,1,0)&2&
(1,0,0,1)&(0,1,1,0)&2\\ \hline (1,0,0,1)&(1,1,1,0)&3&
(0,1,0,1)&(1,1,0,0)&2&(0,1,0,1)&(1,0,1,0)&2\\ \hline
(0,1,0,1)&(0,1,1,0)&2& (0,1,0,1)&(1,1,1,0)&3&
(1,1,0,1)&(1,1,0,0)&3\\ \hline (1,1,0,1)&(0,1,1,0)&3&
(1,1,0,2)&(1,0,1,0)&2& (1,1,0,2)&(1,1,1,0)&3\\ \hline
(0,0,1,1)&(1,1,0,0)&2&(0,0,1,1)&(1,0,1,0)&2&(0,0,1,1)&(0,1,1,0)&2\\ \hline
(0,0,1,1)&(1,1,1,0)&3&(1,0,1,1)&(1,1,0,0)&3&(1,0,1,1)&(0,1,1,0)&3\\ \hline (1,0,1,2)&(1,0,1,0)&2&
(1,0,1,2)&(1,1,1,0)&3& (0,1,1,1)&(1,1,0,0)&3\\ \hline
(0,1,1,1)&(0,1,1,0)&3& (0,1,1,2)&(1,0,1,0)&2&(0,1,1,2)&(1,1,1,0)&3
\\ \hline (1,1,1,2)&(1,1,0,0)&3&
(1,1,1,2)&(1,0,1,0)&3& (1,1,1,2)&(0,1,1,0)&3\\ \hline
(1,1,1,2)&(1,1,1,0)&4&&&&&&\\ \hline
\end{tabular}}
\end{table}

We list all arcs $\vartheta$ with $w(\vartheta)\geq 2$,
$T(\vartheta)\not\in\ S_3$ and $H(\vartheta)\not\in \Gamma$ in Table
6,  where $S_3$ is defined by (\ref{LCL:eq2}).

If $\vartheta_0$ is the arc $(1,1,0,2)\xrightarrow[]{}(1,1,1,1)$ in
Table 6, by Appendix A, $P_0W_0P_3$  has the form as
\begin{equation*}\label{l:equ22}(1,1,0,2)\xrightarrow[]{}(1,1,1,1)
\xrightarrow{(0,1)}(0,0,0,0)\xrightarrow{(1,0)}.\end{equation*}The
vertices $P_j$ for $j\geq3$ are contained in $S_1$ by
Lemma 6 (i). Notice that $(1,1,0,2)\not\in S_1$. Consequently, $W_0$ cannot be closed.

\begin{table}[htbp] \caption{All arcs $\vartheta$ with $w(\vartheta)\geq 2$,
$T(\vartheta)\not\in S_3$ and $H(\vartheta)\not\in \Gamma$}
\center{\begin{tabular}{|c|c|c|c|c|c|c|c|c|} \hline
$T(\vartheta)$&$H(\vartheta)$&$w(\vartheta)$&$T(\vartheta)$&$H(\vartheta)$&
$w(\vartheta)$&$T(\vartheta)$&$H(\vartheta)$&$w(\vartheta)$
\\ \hline (1,1,0,2)&(1,1,1,1)&2&
(1,0,1,2)&(1,1,1,1)&2& (0,1,1,2)&(1,1,1,1)&2\\ \hline
(1,1,1,3)&(1,1,1,1)&2& (1,1,1,1)&(0,0,0,0)&2&(1,1,1,1)&(0,1,0,0)&3\\
\hline (1,1,1,2)&(1,1,0,1)&2& (1,1,1,2)&(0,1,1,1)&2&
(1,1,1,2)&(1,0,0,0)&2
\\\hline (1,1,1,2)&(0,0,1,0)&2&&&&&&\\ \hline
\end{tabular}}
\end{table}

If $\vartheta_0$ is the arc $(1,0,1,2)\xrightarrow[]{}(1,1,1,1)$ in
Table 6, $P_0W_0P_5$  has the form as
\begin{equation*}\label{l:equ23}(1,0,1,2)\xrightarrow[]{}(1,1,1,1)\xrightarrow{(1,1)}(0,1,0,0)
\xrightarrow{(1,-1)}(0,1,0,0)
\xrightarrow{(0,-2)}(0,0,0,0)\xrightarrow{(0,-2)}.\end{equation*}
The vertices $P_j$ for $j\geq 5$ are contained in $S_3$ by Lemma 6
(iii). Therefore, $W_0$ cannot be closed since $(1,0,1,2)\not\in
S_3$. The cases for the arcs $(0,1,1,2)\xrightarrow[]{}(1,1,1,1)$
and $(1,1,1,3)\xrightarrow[]{}(1,1,1,1)$ in Table 6 can be similarly
proven.

If $\vartheta_0$ is the arc $(1,1,1,1)\xrightarrow[]{}(0,0,0,0)$ in
Table 6, then  $P_0W_0P_2$ has the form as
$(1,1,1,1)\xrightarrow[]{}(0,0,0,0)\xrightarrow{(1,1)}$. Thus, all
vertices $P_j$ for $j\geq 2$  are contained in $S_1$ by Lemma 6 (i),
and then $W_0$ cannot be closed since $(1,1,1,1)\not\in S_1$.

If $\vartheta_0$ is the arc $(1,1,1,1)\xrightarrow[]{}(0,1,0,0)$ in
Table 6, $P_0W_0P_4$ has the form as
\[(1,1,1,1)\xrightarrow[]{}(0,1,0,0)\xrightarrow{(1,0)}(0,1,0,0)\xrightarrow{(0,-1)}(0,0,0,0)\xrightarrow{(0,-1)},\]
and the vertices $P_j$ for $j\geq 4$ are contained in $S_2$ by
Lemma 6 (ii). So $W_0$ cannot be closed since
$(1,1,1,1)\not\in S_2$.

If $\vartheta_0$ is the arc $(1,1,1,2)\xrightarrow[]{}(1,1,0,1)$ in
Table 6, $(1,1,0,1)\xrightarrow{(1,1)}$ has an SBS as
\[(1,1,0,1)\xrightarrow{(1,1)}\left\{\begin{array}{l}(0,1,0,0)\xrightarrow{(0,0)}
(0,0,0,0)\xrightarrow{(0,0)}\\(0,1,0,1)\xrightarrow{(0,1)}
\left\{\begin{array}{l}(1,0,0,0)\xrightarrow{(0,1)}(0,0,0,0)\xrightarrow{(0,1)}\\
(0,0,1,0)\xrightarrow{(0,1)}(0,0,0,0)\xrightarrow{(1,1)}\end{array}\right.\end{array}\right.\]
and then  the vertices $P_j$ for $j\geq 4$ are contained in $S_1$ by
Lemma 6 (i). Thus $W_0$ cannot be closed since
$(1,1,1,2)\not\in S_1$.

If $\vartheta_0$ is the arc $(1,1,1,2)\xrightarrow[]{}(0,1,1,1)$ in
Table 6, $(0,1,1,1)\xrightarrow{(1,1)}$ has an SBS as
\[(0,1,1,1)\xrightarrow{(1,1)}\left\{\begin{array}{l}(0,1,0,0)\xrightarrow{(1,0)}(0,1,0,0)
\xrightarrow{(0,-1)}(0,0,0,0)\xrightarrow{(0,-1)}\\
(0,1,0,1)\xrightarrow{(1,1)}\left\{\begin{array}{l}(1,1,0,1)\xrightarrow{(0,1)}\\
(0,1,1,1)\xrightarrow{(0,1)}.\end{array}\right.\end{array}\right.\]
The walks $(0,1,1,1)\xrightarrow{(0,1)}$ and $(1,1,0,1)\xrightarrow{(0,1)}$ have been analyzed in (\ref{NEW1}) and (\ref{NEW2}) in Appendix B, respectively. Thus by Lemma 6, the vertices $P_j$ for $j\geq 1$ are contained in $S_3$. So $W_0$ cannot be closed since $(1,1,1,2)\not\in S_3$.

If $\vartheta_0$ is the arc $(1,1,1,2)\xrightarrow[]{}(1,0,0,0)$ in
Table 6, $P_0W_0P_4$ has the form as
\[(1,1,1,2)\xrightarrow[]{}(1,0,0,0)\xrightarrow{(1,1)}(0,1,0,0)\xrightarrow{(0,0)}(0,0,0,0)\xrightarrow{(0,0)},\]
and the vertices $P_4$ for $j\geq 4$ are contained in $S_1$ by
Lemma 6 (i). So $W_0$ cannot be closed since
$(1,1,1,2)\not\in S_1$.

If $\vartheta_0$ is the arc $(1,1,1,2)\xrightarrow[]{}(0,0,1,0)$ in
Table 6, $P_0W_0P_5$ has the form as
\[(1,1,1,2)\xrightarrow[]{}(0,0,1,0)\xrightarrow{(1,1)}(0,1,0,0)
\xrightarrow{(1,0)}(0,1,0,0)\xrightarrow{(0,-1)}(0,0,0,0)\xrightarrow{(0,-1)}.\]
Thus the vertices $P_j$ for $j\geq 5$ are contained in $S_2$ by
Lemma 6 (ii). So $W_0$ cannot be closed  since $(1,1,1,2)\not\in
S_2$.

The above facts show that if $\vartheta_0$ is any arc in Table 6
then the walk $W_0$ cannot be closed. Suppose that $\vartheta_0$
satisfies $T(\vartheta_0)\in S_3$ and $H(\vartheta_0)\not\in
\Gamma$, i.e., those arcs in Table 7. By Lemma 7, we still have that
the walk $W_0$ cannot be closed for any $\vartheta_0$ given by Table
7. However, by (\ref{vertice}) and the analysis therein, we have
that $W_0$ is closed. This contradiction shows that the assumption
at the beginning of the proof does not hold, and then the proof is
finished. \quad$\blacksquare$

\begin{table}[htbp] \caption{All arcs $\vartheta$ with $w(\vartheta)\geq 2$, $T(\vartheta)\in S_3$ and $H(\vartheta)\not\in \Gamma$} \center{\begin{tabular}{|c|c|c|c|c|c|c|c|c|} \hline $T(\vartheta)$&$H(\vartheta)$&$w(\vartheta)$&$T(\vartheta)$&$H(\vartheta)$&$w(\vartheta)$&$T(\vartheta)$&$H(\vartheta)$&$w(\vartheta)$
\\ \hline (1,0,0,0)&(0,1,0,0)&2& (0,1,0,0)&(0,1,0,0)&2& (0,0,1,0)&(0,1,0,0)&2\\
\hline (1,0,1,1)&(1,0,0,0)&2& (1,0,1,1)&(0,1,0,0)&2& (1,0,1,1)&(0,0,1,0)&2\\ \hline
(1,1,0,1)&(1,0,0,0)&2& (1,1,0,1)&(0,1,0,0)&2& (1,1,0,1)&(0,0,1,0)&2\\
\hline (0,1,1,1)&(1,0,0,0)&2& (0,1,1,1)&(0,1,0,0)&2& (0,1,1,1)&(0,0,1,0)&2\\ \hline
\end{tabular}}
\end{table}

{\bf Remark 3:} In the proof of Proposition 3, we do not distinguish
whether the vertices of the walk $W_0$ are in the set
$\{V_0,V_1,\cdots,V_{m-1}\}$ or not. That is to say, we have proven
that each walk in $\mathcal{P}$ cannot be closed.

{\bf Proposition 4:} For the integer sequence
$\nu_0,\nu_1,\ldots,\nu_{m-1}$ of period $m$, if for any $\nu_i\geq
2$, there exists a positive integer $t\leq m$ such that
$\nu_i+\nu_{i-1}+\cdots+\nu_{i-t+1}\leq  t$, then
$\sum\limits_{i=0}^{m-1}\nu_i\leq m$.

{\bf Proof:}  Let $I=\big\{i\,|\,\nu_i\geq2\big\}$ and $|I|=p$.
Thus, all elements of $I$ can be listed as
$i_1,\,i_2,\,\cdots,i_p$, where $i_1<i_2<\cdots<i_p$. For each
integer $i_j\in I$, there exists a least positive integer $t_j$ such
that
\begin{equation}\label{INQU}\nu_{i_j}+\nu_{i_j-1}+\cdots+\nu_{i_j-t_j+1}\leq
t_j,\end{equation} and let
$N_j=\big\{i_j,\,i_j-1,\,\cdots,\,i_j-t_j+1\big\}$ be a subset of
$\mathbb{Z}_m$. Then the inequality (\ref{INQU}) can be written as
$\sum\limits_{i\in N_j}\nu_i\leq t_j=|N_j|$. Let
$N=\bigcup\limits_{j=1}^pN_j$, and we have that $\nu_i\leq1$ if $i\in
\mathbb{Z}_m\setminus N$.

If $p=1$, $\nu_{i_1}+\nu_{i_1-1}+\cdots+\nu_{i_1-t_1+1}\leq  t_1$.
In this case, the proof follows the fact that other $\nu_j$
satisfies $\nu_j\leq 1$.

If $p\geq2$, we claim that for two integers $j$ and $j'$ with $1\leq
j<j'\leq p$, the sets $N_{j}$ and $N_{j'}$ are disjoint or one
containing another one. Without loss of generality, we take $j=1$ and $j'=2$. Then we have
\begin{equation}\label{inqu(1)}\nu_{i_1}+\nu_{i_1-1}+\cdots+\nu_{i_1-t_1+1}\leq t_1\quad{\rm
and}\quad \,\,\nu_{i_2}+\nu_{i_2-1}+\cdots+\nu_{i_2-t_2+1}\leq
t_2,\end{equation} respectively, where the subscripts are taken
modulo $m$ since the integer sequence has period $m$.

If the above claim is not true, then we have $i_1-t_1+1<i_2-t_2+1\leq i_1<i_2$ and consider the
following sequence
$$\nu_{i_1-t_1+1},\cdots,\nu_{i_2-t_2+1},\nu_{i_2-t_2+2},\cdots,\nu_{i_1},\cdots,\nu_{i_2}.$$
Notice that $t_1$ and $t_2$ are the least positive integers
satisfying (\ref{inqu(1)}). Consequently, we have
$$\nu_{i_2-t_2+1}+\nu_{i_2-t_2+2}+\cdots+\nu_{i_1}>
i_1-i_2+t_2,\,\,\,{\rm
and}\,\,\nu_{i_1+1}+\nu_{i_1+2}+\cdots+\nu_{i_2}> i_2-i_1.$$ This
implies
$$\nu_{i_2-t_2+1}+\nu_{i_2-t_2+2}+\cdots+\nu_{i_1}+\nu_{i_1+1}+\nu_{i_1+2}+\cdots+\nu_{i_2}>t_2,$$
which contradicts with (\ref{inqu(1)}) and then the claim is true. Thus
there exists a subset $J$ of the set $\{1,2,\cdots,p\}$ such
that
$$N=\bigcup\limits_{j\in J}N_{j}\quad{\rm
and}\quad N_{j}\bigcap N_{j'}=\emptyset\quad {\rm for\,\,any\,\,two\,\,different\,\,elements\,\, }
j \,\,{\rm and}\,\,  j'\,\,{\rm of}\,\,  J.$$ Thus
$|N|=\sum\limits_{j\in J}|N_{j}|=\sum\limits_{j\in J}t_{j}$ and we
have that
\[\sum\limits_{i\in N}\nu_i=\sum\limits_{j\in J}\sum\limits_{i\in N_j}\nu_i\leq\sum\limits_{j\in J}t_j=|N|.\]
Therefore, we have
\[\sum\limits_{i=0}^{m-1}\nu_i=\sum\limits_{i\in
\mathbb{Z}_m\setminus N}\nu_i+\sum\limits_{i\in N}\nu_i\leq\sum\limits_{i\in
\mathbb{Z}_m\setminus N}1+|N|=m,\] and this finishes the proof.\quad$\blacksquare$

Propositions 3 and 4 tell us that $M(m;3,13)\leq k$. Furthermore, we
can also prove that the equal sign holds.

{\bf Lemma 8:} (Theorem 14, \cite{HX02}) We have that
\[M(m;2^r+1)=\left\{\begin{array}{ll}m/2,&\mbox{if } m/(r,m)\mbox{ is even},\\
(m-(m,r))/2,&\mbox{if } m/(r,m)\mbox{ is odd}.\end{array}\right.\]

{\bf Proof of Proposition 2:}  By Propositions 3 and 4, we have
$w(\nu)\leq m$ and then by (\ref{Cr4})
$$M(m;3, 13)=\max\left(w(s)-w(a)-w(b)\right)\leq k$$
where the maximum is over all integers $s$, $a$, $b$ such that
$$0\leq s, a, b\leq 2^m-1,\,\,\,\,s\equiv 3a+13b\,({\rm mod}\, 2^m-1),\
\,\,\,a\,\,{\rm or}\,\, b\not\equiv 0\,({\rm mod}\,2^m-1).$$
On the other hand, we have
$M(m;3,13)\geq M(m;3)$ by the definition of $M(m;3,13)$. Applying Lemma 8, we have
$$k=(m-(m,r))/2=M(m;3)\leq M(m;3,13)\leq k.$$
Therefore, we have $M(m;3,13)=k$ and the proof is finished.\quad$\blacksquare$

\section{Concluding Remarks}

For odd $m\geq 5$, a new triple-error-correcting cyclic code of
length $2^m-1$ has been found. It is defined by zeros
$\alpha,\alpha^3$ and $\alpha^{13}$, and the exponents $3$ and $13$
come from the Gold and Kasami-Welch APN power functions,
respectively. To generalize the construction of the code
$\mathcal{C}_{1,3,13}$, one can consider the class of cyclic codes
$\mathcal{C}$ with the dual codes $\mathcal{C}^\perp$ having the
form
$$\begin{array}{c}\mathcal{C}^\bot={\Big\{}{\bf
c}(\epsilon,\gamma,\delta) =\left({\rm Tr}^m_1(\epsilon x+\gamma
f(x)+\delta g(x)\right)_{x\in \mathbb{F}_{2^m}^*}\mid\epsilon ,\,
\gamma,\,\delta\in \mathbb{F}_{2^m} {\Big\}}\end{array}$$ where
$f(x)$ and $g(x)$ are different APN functions from
$\mathbb{F}_{2^m}$ to itself. If the polynomial ${\rm
Tr}^m_1(\epsilon x+\gamma f(x)+\delta g(x))$ in variable $x$ has
algebraic degree greater than $2$, some tools other than the theory
of quadratic forms are possibly needed.

\section*{Appendix A: Some Arcs $\vartheta$ in $\mathbb{D}$}

Appendix A gives all arcs $\vartheta$ with the tail $T(\vartheta)$ in the set
$$\begin{array}{r}\Big\{(0,0,0,0),(0,0,0,1),(1,0,0,0),(1,0,0,1),(0,1,0,0),(0,1,0,1),
(1,1,0,1),(1,1,0,2),(0,0,\\
1,0),(0,0,1,1),
(1,0,1,1),(1,0,1,2),(0,1,1,1),(0,1,1,2),(1,1,1,1),(1,1,1,2),(1,1,1,3)\Big\}\end{array}$$
and head $H(\vartheta)\not\in \Gamma$.

\hspace{-6mm}1. $T(\vartheta)=(0,0,0,0)$.
\[\begin{tabular}{|c|c|c|c|c|c|c|} \hline $H(\vartheta)$&(0,0,0,0)&(0,0,0,1)&(1,0,0,0)&(0,1,0,0)&(0,1,0,1)&(0,0,1,0)\\
\hline $w(\vartheta)$&0&-1&1&1&0&1\\ \hline
\end{tabular}\]
2. $T(\vartheta)=(0,0,0,1)$. \[\begin{tabular}{|c|c|c|c|c|c|c|c|}
\hline $H(\vartheta)$&(0,0,0,2)&(0,0,0,3)&(1,0,0,1)&(1,0,0,2)&(0,1,0,2)&(0,1,0,3)&(1,1,0,1)\\
\hline$w(\vartheta)$&-3&-4&-1&-2&-2&-3&0\\
\hline $H(\vartheta)$&(1,1,0,2)&(0,0,1,1)&(0,0,1,2)&(1,0,1,1)&(0,1,1,1)&(0,1,1,2)&(1,1,1,1)\\ \hline $w(\vartheta)$&-1&-1&-2&0&0&-1&1\\
\hline
\end{tabular}\]
3.  $T(\vartheta)=(1,0,0,0)$. \[\begin{tabular}{|c|c|c|}
\hline $H(\vartheta)$&(0,0,0,0)&(0,1,0,0)\\
\hline $w(\vartheta)$&1&2\\ \hline
\end{tabular}\]
4. $T(\vartheta)=(1,0,0,1)$. \[\begin{tabular}{|c|c|c|c|c|c|}
\hline $H(\vartheta)$&(0,0,0,1)&(0,0,0,2)&(1,0,0,0)&(1,0,0,1)&(0,1,0,1)\\
\hline $w(\vartheta)$&-1&-2&1&0&0\\ \hline $H(\vartheta)$&(0,1,0,2)&(1,1,0,1)&(0,0,1,0)&(0,0,1,1)&(0,1,1,1)\\ \hline
$w(\vartheta)$&-1&1&1&0&1\\ \hline
\end{tabular}\]
5. $T(\vartheta)=(0,1,0,0)$. \[\begin{tabular}{|c|c|c|}
\hline $H(\vartheta)$&(0,0,0,0)&(0,1,0,0)\\
\hline $w(\vartheta)$&1&2\\ \hline
\end{tabular}\]
6.$T(\vartheta)=(0,1,0,1)$. \[\begin{tabular}{|c|c|c|c|c|c|}
\hline $H(\vartheta)$&(0,0,0,1)&(0,0,0,2)&(1,0,0,0)&(1,0,0,1)&(0,1,0,1)\\
\hline $w(\vartheta)$&-1&-2&1&0&0\\ \hline $H(\vartheta)$&(0,1,0,2)&(1,1,0,1)&(0,0,1,0)&(0,0,1,1)&(0,1,1,1)\\ \hline
$w(\vartheta)$&-1&1&1&0&1\\ \hline
\end{tabular}\]
7. $T(\vartheta)=(1,1,0,1)$.
\[\begin{tabular}{|c|c|c|c|c|c|c|}
\hline $H(\vartheta)$&(0,0,0,0)&(0,0,0,1)&(1,0,0,0)&(0,1,0,0)&(0,1,0,1)&(0,0,1,0)\\
\hline $w(\vartheta)$&1&0&2&2&1&2\\ \hline
\end{tabular}\]
8. $T(\vartheta)=(1,1,0,2)$. \[\begin{tabular}{|c|c|c|c|c|c|c|c|}
\hline $H(\vartheta)$&(0,0,0,2)&(0,0,0,3)&(1,0,0,1)&(1,0,0,2)&(0,1,0,2)&(0,1,0,3)&(1,1,0,1)\\
\hline $w(\vartheta)$&-2&-3&0&-1&-1&-2&1\\
\hline $H(\vartheta)$&(1,1,0,2)&(0,0,1,1)&(0,0,1,2)&(1,0,1,1)&(0,1,1,1)&(0,1,1,2)&(1,1,1,1)\\
 \hline $w(\vartheta)$&0&0&-1&1&1&0&2\\ \hline
\end{tabular}\]
9. $T(\vartheta)=(0,0,1,0)$. \[\begin{tabular}{|c|c|c|}
\hline $H(\vartheta)$&(0,0,0,0)&(0,1,0,0)\\
\hline $w(\vartheta)$&1&2\\ \hline
\end{tabular}\]
10. $T(\vartheta)=(0,0,1,1)$.
\[\begin{tabular}{|c|c|c|c|c|c|}
\hline $H(\vartheta)$&(0,0,0,1)&(0,0,0,2)&(1,0,0,0)&(1,0,0,1)&(0,1,0,1)\\
\hline $w(\vartheta)$&-1&-2&1&0&0\\ \hline $H(\vartheta)$&(0,1,0,2)&(1,1,0,1)&(0,0,1,0)&(0,0,1,1)&(0,1,1,1)\\ \hline
$w(\vartheta)$&-1&1&1&0&1\\ \hline
\end{tabular}\]
11. $T(\vartheta)=
(1,0,1,1)$.\[\begin{tabular}{|c|c|c|c|c|c|c|} \hline $H(\vartheta)$&(0,0,0,0)&(0,0,0,1)&(1,0,0,0)&(0,1,0,0)&(0,1,0,1)&(0,0,1,0)\\
\hline $w(\vartheta)$&1&0&2&2&1&2\\ \hline
\end{tabular}\]
12. $T(\vartheta)=(1,0,1,2)$.
\[\begin{tabular}{|c|c|c|c|c|c|c|c|}
\hline $H(\vartheta)$&(0,0,0,2)&(0,0,0,3)&(1,0,0,1)&(1,0,0,2)&(0,1,0,2)&(0,1,0,3)&(1,1,0,1)\\
\hline $w(\vartheta)$&-2&-3&0&-1&-1&-2&1\\
\hline $H(\vartheta)$&(1,1,0,2)&(0,0,1,1)&(0,0,1,2)&(1,0,1,1)&(0,1,1,1)&(0,1,1,2)&(1,1,1,1)\\
\hline $w(\vartheta)$&0&0&-1&1&1&0&2\\\hline
\end{tabular}\]
13. $T(\vartheta)=(0,1,1,1)$.
\[\begin{tabular}{|c|c|c|c|c|c|c|} \hline $H(\vartheta)$&(0,0,0,0)&(0,0,0,1)&(1,0,0,0)&(0,1,0,0)&(0,1,0,1)&(0,0,1,0)\\
\hline $w(\vartheta)$&1&0&2&2&1&2\\ \hline
\end{tabular}\]
14. $T(\vartheta)=(0,1,1,2)$.
\[\begin{tabular}{|c|c|c|c|c|c|c|c|}
\hline $H(\vartheta)$&(0,0,0,2)&(0,0,0,3)&(1,0,0,1)&(1,0,0,2)&(0,1,0,2)&(0,1,0,3)&(1,1,0,1)\\
\hline $w(\vartheta)$&-2&-3&0&-1&-1&-2&1\\
\hline $H(\vartheta)$&(1,1,0,2)&(0,0,1,1)&(0,0,1,2)&(1,0,1,1)&(0,1,1,1)&(0,1,1,2)&(1,1,1,1)\\
\hline $w(\vartheta)$&0&0&-1&1&1&0&2\\
\hline
\end{tabular}\]
15. $T(\vartheta)=(1,1,1,1)$. \[\begin{tabular}{|c|c|c|}
\hline $H(\vartheta)$&(0,0,0,0)&(0,1,0,0)\\
\hline $w(\vartheta)$&2&3\\ \hline
\end{tabular}\]
16. $T(\vartheta)=(1,1,1,2)$.
\[\begin{tabular}{|c|c|c|c|c|c|} \hline $H(\vartheta)$&(0,0,0,1)&(0,0,0,2)&(1,0,0,0)&(1,0,0,1)&(0,1,0,1)\\
\hline $w(\vartheta)$&0&-1&2&1&1\\ \hline $H(\vartheta)$&(0,1,0,2)&(1,1,0,1)&(0,0,1,0)&(0,0,1,1)&(0,1,1,1)\\ \hline
$w(\vartheta)$&0&2&2&1&2\\ \hline
\end{tabular}\]
17. $T(\vartheta)=(1,1,1,3)$. \[\begin{tabular}{|c|c|c|c|c|c|c|c|c|}
\hline $H(\vartheta)$&(0,0,0,3)&(0,0,0,4)&(1,0,0,2)&(1,0,0,3)&(0,1,0,3)&(0,1,0,4)&(1,1,0,2)&(1,1,0,3)\\
\hline $w(\vartheta)$&-3&-4&-1&-2&-2&-3&0&-1\\
\hline $H(\vartheta)$&(0,0,1,2)&(0,0,1,3)&(1,0,1,1)&(1,0,1,2)&(0,1,1,2)&(0,1,1,3)&(1,1,1,1)&(1,1,1,2)\\
\hline $w(\vartheta)$&-1&-2&1&0&0&-1&2&1\\
\hline
\end{tabular}\]

\section*{Appendix B: The Proof of Lemma 6}

\textbf{Proof:} The proofs of Lemma 6 (i) and (ii) are contained in
the proof of Lemma 6 (iii), so we only focus on the proof for (iii).
Furthermore, the proof for the case $P_{j+1}(2)=1$ and $\omega=-2$
is contained in that for the case $P_{j+1}(2)=0$ and $\omega=-2$,
thus we always assume that $P_{j+1}(2)=0$ and $\omega=-2$ in the
sequel. For the same reason as in Example 1, without loss of
generality, we can also assume that the integer $q$ is large enough.

Let $\vartheta_i$ denote the arc with the tail $P_i$ and head
$P_{i+1}$ for each $i\in\{0,1,\cdots,q-1\}$.

Since $P_{j+1}(2)=0$ and $\omega=-2$, by $P_j=(0,0,0,0)$ and
Appendix A, we have
$P_{j+1}\in\big\{(0,0,0,0),(0,0,0,1),(0,0,1,0),(1,0,0,0)\big\}$. If
$P_{j+1}=(0,0,0,0)$, then $w(\vartheta_j)=0$. By a similar analysis
as in (\ref{INQ3}), we have $w(\vartheta_{j+1})\geq -1$.
Consequently, $P_jWP_{j+2}$ has the form as
\begin{equation}\label{NEW5}
(0,0,0,0)\xrightarrow{(0,-2)}
(0,0,0,0)\xrightarrow{(0,\,-1)}\hspace{3.5cm}(\Phi 1).
 \end{equation} For $P_{j+1}\in\big\{(0,0,0,1),(0,0,1,0),(1,0,0,0)\big\}$, by a similar analysis $P_jWP_{j+2}$ has other three possible forms as below.
\begin{displaymath}
(0,0,0,0)\xrightarrow{(0,-2)} \left\{
\begin{array}{ll}
(0,0,0,1)\xrightarrow{(0,\,\,0)}&\hspace{3cm}(\Phi 2)
\\
(0,0,1,0)\xrightarrow{(0,-2)}&\hspace{3cm}(\Phi3)
\\
(1,0,0,0)\xrightarrow{(0,-2)}&\hspace{3cm}(\Phi4).
\\
\end{array}\right.
 \end{displaymath}

In the case ($\Phi1$), $P_{j+2}(2)=0$ and then by Appendix A, we
have
$$P_{j+2}\in\Big\{(0,0,0,0),(0,0,0,1),(0,0,1,0),(1,0,0,0)\Big\}.$$
Since the weights of the arcs with the tail $P_{j+1}$ and heads $(0,0,0,0),\,(0,0,0,1)$,
$(0,0,1,0),(1,0,0,0)$ are $0$, $-1$, $1$, $1$,  respectively, there are four
possible forms for $P_jWP_{j+3}$ as
\begin{displaymath}
(0,0,0,0)\xrightarrow{(0,-2)}(0,0,0,0)\xrightarrow{(0,-1)}\left\{
\begin{array}{ll}
(0,0,0,0)\xrightarrow{(0,\,\,0)}&(\Phi1.1)
\\
(0,0,0,1)\xrightarrow{(0,\,\,1)}  &(\Phi1.2)
\\
(0,0,1,0)\xrightarrow{(0,-1)}&(\Phi1.3)
\\
(1,0,0,0)\xrightarrow{(0,-1)} &(\Phi1.4).
\\
\end{array}\right.
 \end{displaymath}
For the case ($\Phi1.1$), $P_{j+3}(2)=0$ and $w(\vartheta_{j+2})\geq
0$. So $P_{j+3}\in\big\{(0,0,0,0), (1,0,0,0), (0,0,1,0)\big\}$ by
Appendix A. When $P_{j+3}=(0,0,0,0)$, we have $w(\vartheta_{j+2})=0$
and $w(\vartheta_{j+3})\geq 0+1-w(\vartheta_{j+2})=1$. By Example 1,
in the case ($\Phi1.1$) and $P_{j+3}=(0,0,0,0)$, all vertices
$P_{l}$ $(j+1\leq l\leq q)$ occurring in the walk $W$ are contained
in the set $\{(0,0,0,0), (1,0,0,0), (0,1,0,0), (0,0,1,0)\}$, which
is a subset of $S_1$.
When $P_{j+3}=(1,0,0,0)$, by a similar analysis $P_{j+3}WP_{j+5}$ has the form
$$
(1,0,0,0)\xrightarrow{(0,\,0)}(0,0,0,0)\xrightarrow{(0,\,0)}.
$$ When $P_{j+3}=(0,0,1,0)$, $P_{j+3}WP_{j+7}$ has three possible forms
$$
(0,0,1,0)\xrightarrow{(0,\,0)}(0,0,0,0)\xrightarrow{(1,\,0)}\left\{\begin{array}{l}
(0,1,0,1)\xrightarrow{(0,\,1)}\left\{\begin{array}{l}
(0,0,1,0)\xrightarrow{(0,\,1)}
\\
(1,0,0,0)\xrightarrow{(0,\,1)}
\end{array}\right.
\\
(0,1,0,0)\xrightarrow{(0,\,0)}(0,0,0,0)\xrightarrow{(0,\,0)}.
\end{array}\right.
$$

Therefore, for the case ($\Phi1.1$), $(0,0,0,0)\xrightarrow{(0,0)}$ has an SBS as
\begin{equation}\label{LCL:eq5}
(0,0,0,0)\xrightarrow{(0,0)} \hspace{-1mm}\left\{
\begin{array}{lr}
\hspace{-2mm}(0,0,0,0)\xrightarrow{(0,1)} \left\{\begin{array}{l}
(0,0,1,0)
\xrightarrow{(0,1)}(0,0,0,0)\xrightarrow{(1,1)}(0,1,0,0)\\\hspace{3.5cm}\xrightarrow{(0,\,1)}
(0,0,0,0)\xrightarrow{(0,\,1)}
 \\
(1,0,0,0)\xrightarrow{(0,1)}(0,0,0,0)\xrightarrow{(0,1)}
\end{array}\right.
\\
\hspace{-2mm}(0,0,1,0)\xrightarrow{(0,0)}(0,0,0,0)\xrightarrow{(1,0)}\hspace{-1mm}\left\{\begin{array}{l}
(0,1,0,1)\xrightarrow{(0,1)}\left\{\begin{array}{l}
\hspace{-1mm}(0,0,1,0)\xrightarrow{(0,1)}
\\
\hspace{-1mm}(1,0,0,0)\xrightarrow{(0,1)}
\end{array}\right.
\\
(0,1,0,0)\xrightarrow{(0,0)}(0,0,0,0)\xrightarrow{(0,0)}
\end{array}\right.
\\
\hspace{-2mm}(1,0,0,0)\xrightarrow{(0,0)}(0,0,0,0)\xrightarrow{(0,0)},
\\
\end{array}\right.
 \end{equation}
in which all vertices and arcs in $P_{j+2}WP_q$ have occurred for
the case $P_{j+2}=(0,0,0,0)$. Thus, all vertices $P_{l}$ $(j+1\leq
l\leq q)$ occurring in the walk $W$ are contained in the set $S_1$
defined by (\ref{LCL:eq4}). Furthermore, by (\ref{LCL:eq5}), all
walks with the form $(0,0,0,0)\xrightarrow{(\eta,\,\omega)}$ for
$\eta\in\{0,1\}$ and $\omega\in\{0,1\}$ have occurred in
(\ref{LCL:eq5}). This finishes the proof of Lemma 6 (i).

For the case ($\Phi1.2$), by Appendix A, we have $
(0,0,0,1)\xrightarrow{(0,\,1)}O
$, i.e., $q=j+2$ and $P_{j+2}=P_q$.

For the case ($\Phi1.3$), $P_{j+2}WP_{j+6}$ has five possible forms as
\begin{equation}\label{LCL:eq6}
(0,0,1,0)\xrightarrow{(0,-1)}(0,0,0,0)\xrightarrow{(1,-1)}
\left\{\begin{array}{l}
(0,1,0,0)\xrightarrow{(0,-1)}(0,0,0,0)\xrightarrow{(0,-1)}\hspace{5mm}(\Phi1.3.1)
\\
(0,1,0,1)\xrightarrow{(0,0)} \left\{\begin{array}{l}
(0,0,1,0)\xrightarrow{(0,0)}\hspace{5mm}(\Phi1.3.2)
\\
(0,0,1,1)\xrightarrow{(0,1)}\hspace{5mm}(\Phi1.3.3)
\\
(1,0,0,0)\xrightarrow{(0,0)}\hspace{5mm}(\Phi1.3.4)
\\
(1,0,0,1)\xrightarrow{(0,1)}\hspace{5mm}(\Phi1.3.5).
\end{array}\right.
\end{array}\right.
\end{equation}
The walks $(0,0,1,0)\xrightarrow{(0,0)}$ in ($\Phi1.3.2$) and
$(1,0,0,0)\xrightarrow{(0,0)}$ in ($\Phi1.3.4$) have occurred in
(\ref{LCL:eq5}). We need to further analyze the cases ($\Phi1.3.3$)
and ($\Phi1.3.5$). By Appendix A, $(0,0,1,1)\xrightarrow{(0,1)}$ has
an SBS as
\begin{equation}\label{NEW3}
(0,0,1,1)\xrightarrow{(0,1)}\left\{\begin{array}{l}
(0,0,1,0)\xrightarrow{(1,1)}(0,1,0,0)\xrightarrow{(1,0)}(0,1,0,0)\xrightarrow{(0,-1)}(0,0,0,0)\xrightarrow{(0,-1)}
\\
(1,0,0,0)\xrightarrow{(1,1)}(0,1,0,0)\xrightarrow{(0,0)}(0,0,0,0)\xrightarrow{(0,0)}
\end{array}\right.
\end{equation} for the  case
($\Phi1.3.3$) and $(1,0,0,1)\xrightarrow{(0,1)}$ has an SBS as
\begin{equation}\label{NEW4}
(1,0,0,1)\xrightarrow{(0,1)}\left\{\begin{array}{l}
(0,0,1,0)\xrightarrow{(0,1)}(0,0,0,0)\xrightarrow{(1,1)}
\\
(1,0,0,0)\xrightarrow{(0,1)}(0,0,0,0)\xrightarrow{(0,1)}
\end{array}\right.
\end{equation}
for the case ($\Phi1.3.5$).

For the case ($\Phi1.4$), $P_{j+2}WP_{j+4}$ is given by
\begin{displaymath}
(1,0,0,0)\xrightarrow{(0,-1)}(0,0,0,0)\xrightarrow{(0,-1)}.
\end{displaymath}

Notice that the walk $(0,0,0,0)\xrightarrow{(0,-1)}$ in
($\Phi1.3.1$), ($\Phi1.3.3$) and ($\Phi1.4$) has occurred as
$P_{j+1}WP_{{j+2}}$ in (\ref{NEW5}). Therefore, by the above
analysis for ($\Phi1.1$)-($\Phi1.4$) and Lemma 6 (i), in the case
that $P_jWP_{j+2}$ has the form as (\ref{NEW5}), all vertices
$P_{l}$ $(j+1\leq l\leq q)$ occurring in the walk $W$ are contained
in the set $S_2$ defined by (\ref{LCL:eq3}). Furthermore, the walks
$(0,0,0,0)\xrightarrow{(\eta,\,-1)}$ for $\eta\in\{0,1\}$ have
occurred in (\ref{LCL:eq6}). This finishes the proof of Lemma 6
(ii).

For the case ($\Phi2$), $(0,0,0,1)\xrightarrow{(0,0)}(1,0,1,1)$ has an SBS as
\begin{displaymath}
(0,0,0,1)\xrightarrow{(0,0)}(1,0,1,1)\xrightarrow{(0,1)}
\left\{\begin{array}{l} (0,0,0,0)\xrightarrow{(1,1)}
\\
(0,0,1,0)\xrightarrow{(1,0)}(0,1,0,0)\xrightarrow{(1,-1)}(0,1,0,0)\\\hspace{4.5cm}\xrightarrow{(0,-2)}(0,0,0,0)\xrightarrow{(0,-2)}
\\
(1,0,0,0)\xrightarrow{(1,0)}(0,1,0,0)\xrightarrow{(0,-1)}(0,0,0,0)\xrightarrow{(0,-1)}.
\end{array}\right.
\end{displaymath}

For the case ($\Phi3$), $P_{j+1}WP_{j+5}$ has six possible forms as
\begin{equation}\label{Last}
(0,0,1,0)\xrightarrow{(0,-2)}(0,0,0,0)\xrightarrow{(1,-2)}\left\{\begin{array}{l}
(0,1,0,0)\xrightarrow{(0,-2)}(0,0,0,0)\xrightarrow{(0,-2)}\hspace{6mm}\,(\Phi3.1)
\\
(0,1,0,1)\xrightarrow{(0,-1)} \left\{\begin{array}{l}
(0,0,0,1)\xrightarrow{(0,\,1)}\,\hspace{1mm}\,\,(\Phi3.2)
\\
(0,0,1,0)\xrightarrow{(0,-1)}\hspace{1mm}\,(\Phi3.3)
\\
(0,0,1,1)\xrightarrow{(0,\,0)}\hspace{3mm}(\Phi3.4)
\\
(1,0,0,0)\xrightarrow{(0,-1)}\hspace{1mm}\, (\Phi3.5)
\\
(1,0,0,1)\xrightarrow{(0,\,0)}\hspace{2mm}\,\, (\Phi3.6).
\end{array}\right.
\end{array}\right.
\end{equation}
The walk $(0,0,0,1)\xrightarrow{(0,\,1)}$ in  $(\Phi3.2)$ has
occurred as $P_{j+2}WP_{j+3}$ in ($\Phi1.2$). For the case
($\Phi3.3$), since the segment $P_{j+4}WP_{j+5}$ has the form
$(0,0,1,0)\xrightarrow{(0,-1)}$, the segment $P_{j+4}WP_{j+6}$ has
the form
 $(0,0,1,0)\xrightarrow{(0,-1)}(0,0,0,0)\xrightarrow{(1,-1)}$.
 By Lemma 6 (ii), for the cases  $(\Phi3.2)$ and $(\Phi3.3)$,  all vertices in $W$
 are contained in the set $S_2$.

For the case ($\Phi3.4$), $(0,0,1,1)\xrightarrow{(0,0)}$ has an SBS as
\begin{displaymath}
(0,0,1,1)\xrightarrow{(0,0)}\left\{\begin{array}{l}
(0,0,1,0)\xrightarrow{(1,0)}(0,1,0,0)\xrightarrow{(1,-1)}(0,1,0,0)\xrightarrow{(0,-2)}
(0,0,0,0)\xrightarrow{(0,-2)}(\Phi3.4.1)
\\
 (0,0,1,1)\xrightarrow{(1,1)}\left\{\begin{array}{l}(1,1,0,1)\xrightarrow{(1,1)}\hspace{5.5cm}(\Phi3.4.2)\\
 (0,1,1,1)\xrightarrow{(1,1)}\hspace{5.5cm}(\Phi3.4.3)\end{array}\right.

\\
(1,0,0,0)\xrightarrow{(1,0)}(0,1,0,0)\xrightarrow{(0,-1)}(0,0,0,0)\xrightarrow{(0,-1)}\hspace{2.9cm}(\Phi3.4.4)
\\
(1,0,0,1)\xrightarrow{(1,1)}\left\{\begin{array}{l}
(0,1,1,1)\xrightarrow{(0,1)}\hspace{5.5cm}(\Phi3.4.5)
\\
(1,1,0,1)\xrightarrow{(0,1)}\hspace{5.5cm}(\Phi3.4.6).
\end{array}\right.
\end{array}\right.
\end{displaymath}
For the case $(\Phi3.4.2)$, $(0,0,1,1)\xrightarrow{(1,1)}$ has an SBS as
\begin{displaymath}
(0,0,1,1)\xrightarrow{(1,1)}(1,1,0,1)\xrightarrow{(1,1)}\left\{\begin{array}{l}
(0,1,0,0)\xrightarrow{(0,0)}(0,0,0,0)\xrightarrow{(0,0)}
\\
(0,1,0,1)\xrightarrow{(0,1)}
\end{array}\right.
\end{displaymath}
and the walk $(0,1,0,1)\xrightarrow{(0,1)}$ has occurred in (\ref{LCL:eq5}).
For the case $(\Phi3.4.3)$, $P_{j+5}WP_{j+9}$ has three possible forms as
\begin{displaymath}
(0,0,1,1)\xrightarrow{(1,1)}(0,1,1,1)\xrightarrow{(1,1)}\left\{\begin{array}{l}
(0,1,0,0)\xrightarrow{(1,0)}(0,1,0,0)\xrightarrow{(0,-1)}\hspace{1.3cm}(\Phi3.4.3.1)
\\
(0,1,0,1)\xrightarrow{(1,1)}\left\{\begin{array}{l}
(0,1,1,1)\xrightarrow{(0,1)}\hspace{1cm}\,(\Phi3.4.3.2)\\
(1,1,0,1)\xrightarrow{(0,1)}\hspace{1cm}\,(\Phi3.4.3.3).
\end{array}\right.
\end{array}\right.
\end{displaymath}
Since the walk $(0,1,0,0)\xrightarrow{(0,-1)}$ in the case $(\Phi3.4.3.1)$ has occurred in the case $(\Phi1.3.1)$ as (\ref{LCL:eq6}), we need to further analyze the cases $(\Phi3.4.3.2)$ and $(\Phi3.4.3.3)$.  $(0,1,1,1)\xrightarrow{(0,1)}$ has an SBS as
\begin{equation}\label{NEW1}
(0,1,1,1)\xrightarrow{(0,1)}\left\{\begin{array}{l}
(0,0,0,0)\xrightarrow{(1,1)}
\\
(0,0,1,0)\xrightarrow{(1,0)}(0,1,0,0)\xrightarrow{(1,-1)}(0,1,0,0)\xrightarrow{(0,-2)}(0,0,0,0)\xrightarrow{(0,-2)}
\\
(1,0,0,0)\xrightarrow{(1,0)}(0,1,0,0)\xrightarrow{(0,-1)}(0,0,0,0)\xrightarrow{(0,-1)}
\end{array}\right.
\end{equation} for $(\Phi3.4.3.2)$, and $(1,1,0,1)\xrightarrow{(0,1)}$ has an SBS as
\begin{equation}\label{NEW2}
(1,1,0,1)\xrightarrow{(0,1)}\left\{\begin{array}{l}
(0,0,0,0)\xrightarrow{(0,1)}
\\
(0,0,1,0)\xrightarrow{(0,0)}(0,0,0,0)\xrightarrow{(1,0)}
\\
(1,0,0,0)\xrightarrow{(0,0)}(0,0,0,0)\xrightarrow{(0,0)}
\end{array}\right.
\end{equation}
for  $(\Phi3.4.3.3)$.

Notice that the walk $(0,0,0,0)\xrightarrow{(0,-1)}$ in
($\Phi3.4.4$) has occurred in  ($\Phi1$) and the walks
$(0,1,1,1)\xrightarrow{(0,1)}$ in   ($\Phi3.4.5$) and
$(1,1,0,1)\xrightarrow{(0,1)}$ in  ($\Phi3.4.6$) have been analyzed
in (\ref{NEW1}) and (\ref{NEW2}), respectively.

For the case ($\Phi3.5$), $P_{j+4}WP_{j+6}$ has the form as
$(1,0,0,0)\xrightarrow{(0,-1)}(0,0,0,0)\xrightarrow{(0,-1)}$ and
for the case ($\Phi3.6$), $(1,0,0,1)\xrightarrow{(0,0)}$ has an SBS as
\begin{displaymath}
(1,0,0,1)\xrightarrow{(0,0)} \left\{\begin{array}{l}
(0,0,1,0)\xrightarrow{(0,0)}(0,0,0,0)\xrightarrow{(1,0)}
\\
(0,0,1,1)\xrightarrow{(0,1)}
\\
(1,0,0,0)\xrightarrow{(0,0)}(0,0,0,0)\xrightarrow{(0,0)}
\\
(1,0,0,1)\xrightarrow{(0,1)}.
\end{array}\right.
\end{displaymath}
Notice that the walks $(0,0,1,1)\xrightarrow{(0,1)}$ and
$(1,0,0,1)\xrightarrow{(0,1)}$ have been analyzed in (\ref{NEW3})
and (\ref{NEW4}), respectively.

For the case ($\Phi4$), the segment $P_{j+1}WP_{j+3}$ has the form
$(1,0,0,0)\xrightarrow{(0,-2)}(0,0,0,0)\xrightarrow{(0,-2)}$.

Notice that the walk $(0,0,0,0)\xrightarrow{(0,-2)}$ in the cases
($\Phi2$), ($\Phi3.1$), ($\Phi3.4.1$), ($\Phi3.4.3.2$),
($\Phi3.4.5$), and ($\Phi4$)  has occurred as $P_{j}WP_{j+1}$.
Therefore, combining the above analysis for the cases
($\Phi2$)-($\Phi4$) and by Lemma 6 (i), (ii), all vertices $P_{l}$
$(j+1\leq l\leq q)$ occurring in the walk $W$ are contained in the
set $S_3$. The proof for the case $\eta=1$ and $\omega=-2$ is
contained in the analysis of the case ($\Phi3$) in (\ref{Last}).
This finishes the proof of Lemma 6 (iii). \quad$\blacksquare$

\end{document}